\documentclass[aps,prd,superscriptaddress,nofootinbib,tighten,preprint]{revtex4}
\pdfoutput=1
\usepackage[utf8]{inputenc}
\usepackage[english]{babel}
\usepackage{amsmath}
\usepackage{subfigure}
\usepackage{cancel}
\usepackage{color}
\usepackage{amsfonts}
\usepackage{mathtools}
\usepackage{amssymb}
\usepackage{graphicx}
\newcommand{\be}{\begin{equation}}
\newcommand{\ee}{\end{equation}}
\newcommand{\bea}{\begin{eqnarray}}
\newcommand{\eea}{\end{eqnarray}}

\newcommand{\umt}{{\rm U(1)}_{\rm L_{\mu}-L_{\tau}}}
\newcommand{\ubl}{{\rm U(1)}_{\rm B-L}}
\newcommand{\ublmt}{{{\rm U(1)}_{\rm B-L} \times \rm U(1)}_{\rm L_{\mu}-L_{\tau}}}
\newcommand{\vmt}{v_{\mu \tau}}
\newcommand{\gmt}{g_{\mu \tau}}
\newcommand{\gbl}{g_{BL}}
\newcommand{\zmt}{Z_{\mu \tau}}
\newcommand{\zbl}{Z_{BL}}

\newcommand{\smgauge}{{\rm SU}(3)_c\times{\rm SU}(2)_{\rm L}\times {\rm U}(1)_{\rm Y}}
\newcommand{\nn}{\nonumber}

\catcode`@=12

\begin{document}

\title{Inverse seesaw and dark matter in a
gauged ${\rm B-L}$ extension with flavour symmetry}

\author{Anirban Biswas}
\email{anirban.biswas.sinp@gmail.com}
\affiliation{Indian Institute of Technology Guwahati,
Assam, 781039, India}
\author{Sandhya Choubey}
\email{sandhya@hri.res.in}
\affiliation{Harish-Chandra Research Institute, HBNI,
Chhatnag Road, Jhunsi, Allahabad 211 019, India}
\affiliation{Department of Theoretical Physics, School of
Engineering Sciences, KTH Royal Institute of Technology,
AlbaNova University Center, 106 91 Stockholm, Sweden}
\author{Sarif Khan}
\email{sarifkhan@hri.res.in}
\affiliation{Harish-Chandra Research Institute, HBNI,
Chhatnag Road, Jhunsi, Allahabad 211 019, India}

\begin{abstract}
We propose a model which generates neutrino masses by
the inverse seesaw mechanism, provides a viable dark matter
candidate and explains the muon ($g-2$) anomaly. The
Standard Model (SM) gauge group is extended with a gauged
U(1)$_{\rm B-L}$ as well as a gauged U(1)$_{\rm L_{\mu} - L_{\tau}}$. 
While U(1)$_{\rm L_{\mu} - L_{\tau}}$ is anomaly free, the anomaly
introduced by U(1)$_{\rm B-L}$ is cancelled between the six
SM singlet fermions introduced for the inverse seesaw mechanism
and four additional chiral fermions introduced in this model.
After spontaneous symmetry breaking the four chiral fermionic
degrees of freedom combine to give two Dirac states. The
lightest Dirac fermion becomes stable and hence the 
dark matter candidate. We focus on the region of the
parameter space where the dark matter annihilates to the 
right-handed neutrinos, relating the dark matter sector
with the neutrino sector. 
The U(1)$_{\rm L_{\mu} - L_{\tau}}$ gauge symmetry provides
a flavour structure to the inverse seesaw framework, successfully
explaining the observed neutrino masses and mixings. We study
the model parameters in the light of neutrino oscillation
data and find correlation between them. Values of some of
the model parameters are shown to be mutually exclusive
between normal and inverted ordering of the neutrino mass
eigenstates. Moreover, the muon ($g-2$) anomaly
can be explained by the additional contribution
arising from U(1)$_{\rm L_{\mu} - L_{\tau}}$ gauge boson.
\end{abstract}
\maketitle
\section{Introduction}
\label{Intro}
Even though the Standard Model (SM) of elementary particles
has been very successful in describing the nature around us, 
it is unable to explain all the observed phenomena. 
The main puzzles which SM can not explain are the
presence of neutrino masses and mixings observed in the
neutrino oscillation data \cite{Cowan:1992xc, Fukuda:1998mi,
Ahmad:2002jz, Eguchi:2002dm, An:2015nua, RENO:2015ksa,
Abe:2014bwa, Abe:2015awa, Salzgeber:2015gua, Adamson:2016tbq,
Adamson:2016xxw}, the matter antimatter
asymmetry of the universe \cite{Sakharov:1967dj, Buchmuller:1992qc,
Buchmuller:1996pa, Dulaney:2010dj} and the presence of
non baryonic matter, also called the dark matter (DM)
\cite{Sofue:2000jx, Bartelmann:1999yn, Clowe:2003tk,
Biviano:1996bg, Kahlhoefer:2013dca, Harvey:2015hha,
Hinshaw:2012aka, Ade:2015xua}. 



In this work, we study in detail two of the above
mentioned puzzles, {\it viz.}, the presence of masses and mixings
of neutrinos and DM, and propose an extension of the 
SM that can successfully describe the two observed phenomena. 
One of the simplest ways to explain the tiny neutrino masses naturally
is to extend the SM by an additional
U(1)$_{\rm B-L}$ gauge symmetry which dictates the
introduction of three RH-neutrinos due to anomaly
cancellation.
However, the price one pays in this model
is that in order to get the correct neutrino masses
either the RH neutrinos masses have to be of the order
of GUT scale or the corresponding Yukawa couplings have to
be kept small giving a very small active-sterile
mixing angle ($\frac{m_D}{M_R} \sim 10^{-6}$)
\cite{Khalil:2007dr, FileviezPerez:2010ek, Kikuchi:2008xu,
Fonseca:2011vn, Biswas:2017tce, Biswas:2016yan}. Hence, it
is difficult to test this model at the collider. The other
draw-back of the U(1)$_{\rm B-L}$ extension with three RH
neutrinos is that it has no prediction for neutrino mixing angle. 
The first concern regarding testability of the seesaw
mechanism can be addressed using the so called inverse
seesaw mechanism (ISS) \cite{Mohapatra:1986aw, Mohapatra:1986bd,
Kang:2006sn, An:2011uq, BhupalDev:2012ru, Banerjee:2015gca,
Dev:2009aw,Das:2012ze, Das:2014jxa, Mondal:2016kof, Banerjee:2013fga,
Mondal:2012jv, Matsumoto:2010zg, Humbert:2015epa, Humbert:2015yva,
Ibarra:2011xn, Datta:1992qw, Huitu:2008gf, Khalil:2015wua, Abbas:2015zna,
Elsayed:2011de, Khalil:2015naa, Abdallah:2015uba, Arganda:2015ija},
wherein one can have lower the mass scale of RH neutrinos
while keeping the Yukawa couplings large. The second concern
regarding correct prediction of the neutrino mixing pattern
can be addressed by extending the model further by a flavour symmetry.
Although, in principle one could add a discrete horizontal symmetry
\footnote{If the introduced discrete flavour symmetry is broken
to explain neutrino mixing angles, it leads to the complication
associated with domain walls.}, we choose in this work
U(1)$_{\rm L_{\mu} - L_{\tau}}$ gauged flavour symmetry 
to explain the mixing angles of the
light neutrinos \cite{He:1990pn, He:1991qd, Ma:2001md,
Xing:2015fdg, Biswas:2016yan, Biswas:2016yjr, Biswas:2017ait}.
Moreover, this has an added benefit as the extra neutral gauge
boson $Z_{\mu\tau}$ coming from the  
U(1)$_{\rm L_{\mu} - L_{\tau}}$ gauged flavour symmetry
can provide additional contribution to the muon magnetic
moment and thus explain the muon ($g-2$) data
\cite{Bennett:2004pv, Jegerlehner:2009ry, Agashe:2014kda}.
Furthermore, the U(1)$_{\rm L_{\mu} - L_{\tau}}$ symmetry 
reduces the number of free parameters in the neutrino sector
and provides a very peculiar structure to the neutrino mass matrices.
Hence, we expect sharp correlations among the parameters that
satisfy the neutrino oscillation data. We have analysed the model 
and present results both for the normal hierarchy (NH) where 
the third neutrino mass eigenstate is taken as the heaviest, as well
as the inverted hierarchy (NH) where the third neutrino mass
eigenstate is assumed to be the lightest one. 

Since we introduce two additional gauge symmetries
U(1)$_{\rm B-L}$  and U(1)$_{\rm L_{\mu} - L_{\tau}}$, 
we need to choose the U(1) gauge group charges of all fermions
in such a way that the chiral anomalies cancel consistently.
While U(1)$_{\rm L_{\mu} - L_{\tau}}$ is known to be anomaly
free \cite{He:1990pn, He:1991qd, Ma:2001md, Xing:2015fdg},
the gauged $\ubl$ extension of the SM
is anomalous and hence extra fermionic degrees 
of freedom are needed to make the
theory anomaly free. In our model, we require
six additional fermion singlets of SM for the
neutrino mass generation via ISS mechanism.\,\,In addition,
we also need a viable DM candidate. 
All three requirements are met consistently by introducing 
three usual RH neutrinos $N_{\alpha}$s
with ${\rm B-L}$ charge $-1$, three additional RH
neutrinos $N^{\prime}_{\alpha}$s having opposite ${\rm B-L}$ charge
$+1$ and four chiral fermions $\xi_L$, $\eta_L$, ${\chi_1}_R$ and
${\chi_2}_R$ with fractional ${\rm B-L}$ and
${\rm L_{\mu} -L_{\tau}}$ charges, such as $\frac{4}{3}$,
$\frac{1}{3}$, $-\frac{2}{3}$, $-\frac{2}{3}$
and $\frac{1}{3}$, $\frac{4}{3}$, $\frac{1}{3}$,
$\frac{4}{3}$, respectively.\,\,The above charge assignment cancels out all the ${\rm B-L}$
and ${\rm L_{\mu} -L_{\tau}}$ anomalies consistently while
allowing all the necessary terms in the Lagrangian for
the ISS mechanism. Besides after the spontaneous symmetry
breaking, the lightest of the two massive Dirac states created out
of the four chiral fermions $\xi_L$, $\eta_L$, ${\chi_1}_R$ and
${\chi_2}_R$ is stable in this model and becomes
the DM candidate of the universe. 
The anomaly cancellation by assigning fractional ${\rm B-L}$
charges to the additional fermions has been proposed
in \cite{Patra:2016ofq}. Thereafter, both thermal as
well as non-thermal dark matter phenomenologies in this {\it new}
${\rm B-L}$ model have been studied in
\cite{Patra:2016ofq, Biswas:2016iyh, Nanda:2017bmi} . In this
work, we have mainly concentrated on that portion of the parameter
space where our DM candidate dominantly annihilates to
heavy as well as light neutrinos, which enables
us to find a strong correlation between the neutrino mass generation
and DM freeze-out mechanism.

Rest of the work is arranged in the following manner. In section \ref{model}
we describe the model. In section \ref{result} we present our result
for neutrino and dark matter sector. Finally, in section \ref{conclusion}
we conclude.     
\section{Model}
\label{model}
In this section, we will describe the present model briefly.
In the present work, we have extended all three sectors of
SM, namely the gauge sector, the fermionic sector as well as the scalar sector. 
The additional particle content of the present model is shown in Table \ref{tab1}. 
The extensions in the fermionic sector are necessary for
anomaly cancellation while that in the scalar sector is required
for spontaneous breaking of the additional symmetries and 
the mass generation of the extra fermionic fields. 
We have extended the gauge sector by imposing two additional local U(1)
gauge symmetries. Thus, the complete gauge group under which
the Lagrangian remains invariant before 
spontaneous symmetry breaking is $\smgauge\times\ublmt$.
The addition of these extra U(1) symmetries introduce new 
anomalies in the theory. We know that the SM extended
by the gauged $\umt$ symmetry is anomaly free
\cite{He:1990pn, He:1991qd, Ma:2001md, Xing:2015fdg}. Here,
actually anomaly cancels between second and third generations
of leptons. On the other hand, gauged $\ubl$ extension of the SM
is anomalous and one thus needs to add extra fermionic degrees
of freedom to the particle content of the SM to make the
theory anomaly free. The minimal way to cancel ${\rm B-L}$
anomaly is by adding three RH neutrinos having
${\rm B-L}$ charge $-1$. However, in the present work
our motivation is to study both inverse seesaw
mechanism \cite{Khalil:2010iu, Zhou:2012ds, Law:2013gma, El-Zant:2013nta}
and WIMP dark matter within a complete
model. Therefore, besides the three usual RH neutrinos ($N_{\alpha}$)
with ${\rm B-L}$ charge $-1$, we have introduced three more RH
neutrinos ($N^{\prime}_{\alpha}$) having opposite ${\rm B-L}$ charge
i.e. $+1$. As we have already mentioned earlier that
gauged ${\rm L_{\mu}-L_{\tau}}$ extension of the SM is
non-anomalous, hence we have assigned the ${\rm L_{\mu}-L_{\tau}}$
charges of RH neutrinos in a such way that their contribution to axial
vector anomaly \cite{Adler:1969gk, Bardeen:1969md} and mixed gravitational-gauge
anomaly \cite{Delbourgo:1972xb, Eguchi:1976db} cancels among themselves.
On the other hand, the ${\rm B-L}$ anomaly cancellation requires
more fermionic states with appropriate ${\rm B-L}$ charges.
We thus add four chiral fermions $\xi_L$, $\eta_L$, ${\chi_1}_R$ and
${\chi_2}_R$ with fractional ${\rm B-L}$ charges $\frac{4}{3}$,
$\frac{1}{3}$, $-\frac{2}{3}$, $-\frac{2}{3}$. The
${\rm L_{\mu}-L_{\tau}}$ charges of these chiral fermionic states
cannot be zero due to the presence of mixed anomaly
like $\ubl\,\left[\umt\right]^2$. We will see below that their
charges will be fixed from the anomaly cancellation conditions
coming from $\ubl\,\left[\umt\right]^2$, $\left[\ubl\right]^2\,\umt$,
$\left[\umt\right]^3$ and Gravity$^2\,\umt$, respectively.
Let us assume the ${\rm L_{\mu}-L_{\tau}}$ charges of $\xi_L$,
$\eta_L$, ${\chi_1}_R$ and ${\chi_2}_R$ are $a,\,b,\,c$ and $d$. With
this charge assignment we will check various anomaly cancellation
conditions for both $\ubl$ and $\umt$ gauge groups.
\begin{eqnarray}
\left[\ubl\right]^3 &:& \left[\ubl\right]^3_{\rm SM} +
\left[\ubl\right]^3_{\rm new physics}\,, \nonumber\\
&=& -3 + \left[\underbrace{-3\times (-1)^3 - 3\times (+1)^3}_{\text{contributions from RH neutrinos}} +
\underbrace{\left(\dfrac{4}{3}\right)^3 + \left(\dfrac{1}{3}\right)^3
- \left(-\dfrac{2}{3}\right)^3
- \left(-\dfrac{2}{3}\right)^3}_{\text{contributions from exotic chiral fermions}}
\right]\,,\,\nonumber\\
&=& 0\,,
\end{eqnarray}
and
\begin{eqnarray}
\left[{\rm Gravity}^2\times\ubl\right] &:&
\left[{\rm Gravity}^2\times\ubl\right]_{\rm SM} +
\left[{\rm Gravity}^2\times\ubl\right]_{\rm new physics}\,, \nonumber\\
&=& -3 + \left[\underbrace{-3\times (-1) -
3\times (+1)}_{\text{contributions from RH neutrinos}} + \overbrace{\dfrac{4}{3}
+ \dfrac{1}{3} - \left(-\dfrac{2}{3}\right)
- \left(-\dfrac{2}{3}\right)}^{\text{contributions from exotic chiral fermions}} \right]\,,\,\nonumber\\
&=& 0\,.
\end{eqnarray}
Similarly, for $\umt$:
\begin{eqnarray}
\left[\umt\right]^3&:& \left[\umt\right]^3_{\rm SM} +
\left[\umt\right]^3_{\rm new physics}\,, \nonumber\\
&=&0 +\left[- \underbrace{\left(1\times(1)^3 +1\times(-1)^3+
1\times(-1)^3 +1\times(1)^3\right)}_{\text{for RH neutrinos}} +
\overbrace{a^3 + b^3 -c^3 -d^3}^{\text{for exotic chiral fermions}}\right]\,, \nonumber \\
&=& a^3 + b^3 -c^3 -d^3\,\,
\label{ano_eq1}
\end{eqnarray}
and 
\begin{eqnarray}
\left[\text{Gravity}^2 \umt \right]&:& \left[\umt\right]_{\rm SM} +
\left[\umt\right]_{\rm new physics}\,, \nonumber\\
&=&0 + \left[- \underbrace{\left(1\times(1) +1\times(-1)+
1\times(-1) +1\times(1)\right)}_{\text{for RH neutrinos}} +
\overbrace{a + b -c -d}^{\text{for exotic chiral fermions}}\right]\,, \nonumber \\
&=& a+b-c-d\,,
\label{ano_eq2}
\end{eqnarray}
Now, for $\ubl$ and $\umt$ mixed anomalies:
\begin{eqnarray}
\left[\ubl\right]^2 \umt &:& \left[\left[\ubl\right]^2 \umt\right]_{\rm SM} +
\left[\left[\ubl\right]^2 \umt\right]_{\rm new physics}\,, \nonumber\\
&=& 0 + \left[\left(\dfrac{4}{3}\right)^2 a
+\left(\dfrac{1}{3}\right)^2 b -\left(-\dfrac{2}{3}\right)^2 c
-\left(-\dfrac{2}{3}\right)^2 d\right]\,, 
\nonumber \\
&=& \left(\frac{4}{3}\right)^2\,a + \left(\frac{1}{3}\right)^2\, b
- \left(\frac{2}{3}\right)^2\,c - \left(\frac{2}{3}\right)^2\, d\,,
\label{ano_eq3}
\end{eqnarray}

\begin{eqnarray}
\ubl\left[\umt\right]^2 &:& \left[\ubl\left[\umt\right]^2\right]_{\rm SM} +
\left[\ubl\left[\umt\right]^2\right]_{\rm new physics}\,, \nonumber\\
&=& -2 + \left[\dfrac{4}{3}\,a^2
+\dfrac{1}{3}\,b^2 -\left(-\dfrac{2}{3}\right) c^2
-\left(-\dfrac{2}{3}\right) d^2\right]\,, 
\nonumber \\
&=& -2 + \dfrac{4}{3}\,a^2
+\dfrac{1}{3}\,b^2 + \dfrac{2}{3} c^2
+ \dfrac{2}{3} d^2\,,
\label{ano_eq4}
\end{eqnarray}

Equating Eqs. (\ref{ano_eq1}-\ref{ano_eq4}) to zero we get
four constraint equations, which are
\begin{eqnarray}
&& a^3 + b^3 -c^3 -d^3 = 0\,,\nonumber \\
&& a+b-c-d = 0\,, \nn \\
&& 16\,a + b - 4\,c - 4\,d = 0\,, \nn \\
&& 4 a^2+b^2+2\,c^2+2\,d^2 = 6
\end{eqnarray}
After solving the above four equations simultaneously,
we get the following set of solutions for the
charges of exotic chiral fermions:
\begin{eqnarray}
(a,b,c,d) &=& (\pm 1/3, \pm 4/3, \pm 4/3, \pm 1/3 )\,,\,\, 
(\pm 1/3, \pm 4/3, \pm 1/3, \pm 4/3 )\,,\, \nonumber \\
&&( 0, 0, \pm \sqrt{3/2}, \mp \sqrt{3/2})\,.
\end{eqnarray}
In this work, we have adopted $(a,b,c,d) = (1/3,4/3,1/3,4/3)$.
Furthermore, in addition to the usual SM Higgs doublet, we have
also introduced three singlet scalars $\phi_{i}$ ($i=1$ to 3) with
properly chosen $\ubl$ and $\umt$ charges. Among these scalars,
$\phi_1$ and $\phi_2$ are required to generate masses
for the chiral fermions ($\xi_L$, $\eta_L$, $\chi_{1R}$ and $\chi_{2R}$)
in a gauge invariant manner while the remaining one, $\phi_3$, 
is important for writing the interaction terms between $N_{\alpha}$
and $N^{\prime}_{\beta}$. The later ones are required for successful
implementation of inverse seesaw mechanism within the present scenario. 
In Table \ref{tab1}, we have listed all the new particles introduced
for the present model and their corresponding charges under $\ubl$
and $\umt$ symmetry groups.
\begin{center}
\begin{table}[h!]
\begin{tabular}{||c|c|c|c||}
\hline
\hline
\begin{tabular}{c}
   \\
   \\
   \\
        Fermions\\
    \\
    \\
    \\
    \hline
    \\
    Scalars\\
\end{tabular}
&

\begin{tabular}{c}
    \multicolumn{1}{c}{Fields}\\ 
    \hline
    ($N_{e}$, $N_{e}^{\prime}$)\\ 
    \hline
    ($N_{\mu}$, $N_{\mu}^{\prime}$)\\ 
    \hline
    ($N_{\tau}$, $N_{\tau}^{\prime}$)\\ 
    \hline
    $\xi_{L}$\\ 
    \hline
    $\eta_{L}$\\
    \hline
    $\chi_{1R}$\\
    \hline
    $\chi_{2R}$\\
    \hline
    $\phi_{1}$\\
    \hline
    $\phi_{2}$\\
    \hline
    $\phi_{3}$\\
\end{tabular}
&
\begin{tabular}{c}
    \multicolumn{1}{c}{$\ubl$}\\
    \hline
    ($-1,+1$)\\
    \hline
    ($-1,+1$)\\
    \hline
    ($-1,+1$)\\
    \hline
    $4/3$\\
    \hline
    1/3\\
    \hline
    $-2/3$\\
    \hline
    $-2/3$\\
    \hline
    $1$\\
    \hline
    $2$\\
    \hline
    $0$\\
\end{tabular}
&
\begin{tabular}{c}
    \multicolumn{1}{c}{$\umt$}\\
    \hline
    ($0,0$)\\
    \hline
    ($+1,-1$)\\
    \hline
    ($-1,+1$)\\
    \hline
    $1/3$\\
    \hline
    $4/3$\\
    \hline
    $1/3$\\
	\hline
    $4/3$\\
    \hline
    $0$\\
    \hline
    $0$\\
    \hline
    $1$\\
  
\end{tabular}\\
\hline
\hline
\end{tabular}
\caption{BSM particles and their corresponding charges under the $\ubl$
and $\umt$ gauge group where all of them are singlet under SM gauge group
($\smgauge$).}
\label{tab1}
\end{table}
\end{center}
 
The complete gauge invariant Lagrangian for the
present model is thus given by,
\begin{eqnarray}
\mathcal{L} &=& \mathcal{L}_{SM} + \mathcal{L}_{N} + 
\sum_{i=1}^{3}\,\left(D^i_{\mu} \phi_i\,\right)^{\dagger}
\left({D^i}^{\mu} \phi_i\,\right)
-\mathcal{V}(\phi_h,\phi_1,\phi_2,\phi_3) + \mathcal{L}_{DM} \nonumber \\
&-&\frac{1}{4} F^{BL}_{\rho \sigma}{F^{BL}}^{\rho \sigma} 
-\frac{1}{4} F^{\mu\tau}_{\rho \sigma}{F^{\mu\tau}}^{\rho \sigma}\,,
\label{Lblmt}
\end{eqnarray}
where $\mathcal{L}_{SM}$ is the SM Lagrangian and $\mathcal{L}_{N}$
represents the Lagrangian for the extended neutrino sector. The
extended scalar sector Lagrangian is denoted by the third and fourth
term of the above equation while the dark sector Lagrangian, containing
the interaction terms of chiral fermions, is defined by the
term $\mathcal{L}_{DM}$. Finally, last two terms are
the kinetic terms for the ${\rm B-L}$ and ${\rm L_{\mu}-L_{\tau}}$
gauge bosons in terms of the respective field strength tensor.
Below we have discussed in detail about all the parts of the
Lagrangian written in Eq.~(\ref{Lblmt}).   

\subsection{Extended Scalar Sector}
\label{Ex-scalar-sect}

The Lagrangian for the extended scalar sector of the present
model is given in Eq.~(\ref{Lblmt}). The potential
$\mathcal{V}(\phi_{h},\phi_{1},\phi_{2},\phi_{3})$ appearing
in Eq.~(\ref{Lblmt}) contains all types of interaction terms
among the scalar fields, which are allowed by $\smgauge\times\ubl\times\umt$
gauge symmetries. Therefore, the expression of
$\mathcal{V}(\phi_{h},\phi_{1},\phi_{2},\phi_{3})$ can be written as 
\begin{eqnarray}
\mathcal{V}(\phi_{h},\phi_{1},\phi_{2},\phi_{3}) &=& \mu_{h}^{2} 
(\phi_{h}^{\dagger}\phi_{h}) + \lambda_{h} (\phi_{h}^{\dagger}\phi_{h})^{2}
+ \sum_{i=1}^{3} \left(\mu_{i}^{2}(\phi_{i}^{\dagger}\phi_{i}) +
\lambda_{i} (\phi_{i}^{\dagger}\phi_{i})^{2} \right) \nn \\ &&
+ \tilde{\mu} (\phi_{2} \phi_{1}^{\dagger\,^2} 
+ \phi_{2}^{\dagger} \phi_{1}^{2})
+ \sum_{i=1}^{3} \rho_{i} (\phi_{h}^{\dagger} \phi_{h})
(\phi_{i}^{\dagger} \phi_{i}) \nn \\ &&
+ \sum_{i,j =1,j>i}^{3} \lambda_{ij} (\phi_{i}^{\dagger} \phi_{i})
(\phi_{j}^{\dagger} \phi_{j})\,.
\label{scalar-potential}
\end{eqnarray}

As we want the gauge symmetry to be spontaneously broken
to SU(3)$_c\times{\rm U(1)}_{\rm em}$ {\it i.e.},
$\smgauge\times\ubl\times\umt \xRightarrow[\langle\phi_i\rangle\,
=\,v_i]{\langle \phi_h \rangle\,=\,v/\sqrt{2}}
{\rm SU(3)}_c\times{\rm U(1)}_{\rm em}$, hence the coefficients
of all the quadratic terms must be negative, {\it i.e.}, $\mu_h^2<0$
and $\mu_i^2<0$ (for $i=1$ to 3). It is now well established that the
only spin zero resonance observed in the LHC has properties very
similar to the SM Higgs boson. This actually tells us that the
mixing between Higgs doublet ($\phi_{h}$) and the other scalars ($\phi_i$)
will be inevitably small. Therefore in the current work,
just for the sake of simplicity, we take the mixing angles
between the SM-like Higgs boson and the other non-standard scalars
as equal to zero, {\it i.e.}, $\rho_i=0$ (for $i=1,2, 3$).   
On the other hand, we need to consider mixing among the remaining three
(BSM) scalars $\phi_1$, $\phi_2$ and $\phi_3$. Handling the
mixing among three scalars simultaneously is a tedious job,
hence we will take the vacuum
expectation value (VEV) of $\phi_1$
large enough and correspondingly the mass of the neutral
component such that it will have negligible effect on the
relic density for the mass range
we are considering in the current work.
Moreover, we particularly focus on the parameter space of the model
where DM and heavy neutrinos as well as light neutrinos
are one to one related which means that a reasonable
portion of DM annihilate to these heavy and light neutrinos. Therefore,
we will consider only mixing between $\phi_2$ and $\phi_3$
scalars, {\it i.e.}, we take $\lambda_{23} \neq 0$ while all the other mixing terms we will
neglect to focus on the above mentioned parameter space. 
Although we take all the other quartic mixing terms except $\lambda_{23}$ to be 
equal to zero, but mixing term will be generated due to
the presence of the trilinear coupling $\tilde{\mu}$.
Since $\tilde{\mu}$ is a dimensionful quantity,
its magnitude can be of any order. In the current, work we have
adopted small value for $\tilde{\mu}$ so that the mixing
term generated due to the trilinear term can be
safely neglected.  

After getting VEV, the neutral components of all the scalars
take the following form,
\begin{eqnarray}
\phi_{h}^{} = \frac{v + h + i\, G_h}
{\sqrt{2}},\,\,\, \phi_{j}^{} = 
\frac{v_{j} + H_j + i\,A_j}{\sqrt{2}}\,\,, 
\end{eqnarray}
where $j=1$ to 3 and $G_h$ represents the massless Goldstone
boson which gives mass to the SM neutral gauge boson $Z$.
On the other hand, $A_j$ corresponds to the
CP odd neutral component of the singlet scalar field $\phi_j$.
Among them, $A_3$ and one linear combination of $A_1$, $A_2$ will be massless
as those are responsible for the mass generation of the ${\rm L_{\mu} - L_{\tau}}$
and ${\rm B-L}$ gauge bosons, respectively. The mass of other CP odd
state ($A$), which is orthogonal to the massless state, is given by
\begin{eqnarray}
M^2_A = -\dfrac{\tilde{\mu}\,v_2}{\sqrt{2}}\left(r^2_{{\it vev}}+4\right)\,,
\end{eqnarray} 
where $r_{{\it vev}}= \dfrac{v_1}{v_2}$, the ratio between the two VEVs
$v_1$, $v_2$ and since $M^2_{A}>0$, this implies $\tilde{\mu}<0$.
Moreover, as both the singlet scalars $\phi_1$, $\phi_2$
have nonzero ${\rm B-L}$ charges, hence, they both contribute
to the mass of $\ubl$ gauge boson $\zbl$ and it has the
following form,
\begin{eqnarray}
M^2_{Z_{BL}} &=& g^2_{BL}\left(v^2_1 + 4 v^2_2 \right)\,, \nn \\
v^2_2 &=& \frac{M^2_{Z_{BL}}}{g^2_{BL}\, (r^2_{{\it vev}} + 4)}\,.
\label{v2}
\end{eqnarray}  
The remaining singlet scalar $\phi_3$ is the only member in the scalar
sector which has nonzero $L_{\mu} - L_{\tau}$ charge
hence the mass of the gauge boson $\zmt$ appears when $\phi_3$
gets a VEV, {\it i.e.},
\begin{eqnarray}
M^2_{Z_{\mu \tau}} = g^2_{\mu \tau} v^2_{\mu \tau}\,,
\label{massZmt}
\end{eqnarray}
where we have denoted $v_{3}$ by $v_{\mu\tau}$. 
As we have considered only the mixing between $\phi_2$ and $\phi_3$,
hence the mass matrix with respect to the basis ($H_2$, $H_3$) takes
the following form,
\begin{eqnarray}
M^2_{H_2 H_3} = 
\left(\begin{array}{cc}
2 \lambda_2 v^2_2 ~~&~~ \lambda_{23} v_2 v_3 \\
\lambda_{23} v_2 v_3 ~~&~~ 2 \lambda_3 v^2_3\\
\end{array}\right) \,.
\label{mncomplex}
\end{eqnarray}
By diagonalizing the above mass matrix one can easily obtain
mass basis (physical states) from the gauge basis through an orthogonal
transformation by the mixing angle $\beta$ in the
following manner,
\begin{eqnarray}
h_2 = \cos \beta \,H_2 - \sin \beta \, H_3\,, \\ \nn
h_3 = \sin \beta\, H_2 + \cos \beta \,H_3\,.
\end{eqnarray}
Now, we can write down the quartic couplings related to
$\phi_2$, $\phi_3$ in terms of the masses $M_{h_2}$, $M_{h_3}$
and the mixing angle $\beta$ and have the following form, 
 \begin{eqnarray}
\lambda_2 &=& \frac{(M_{h_2}^2 + M_{h_3}^2) + (M_{h_2}^2 - M_{h_3}^2) \cos 2 \beta }
{4 v_2^{2}}\,, \nn \\
\lambda_3 &=& \frac{(M_{h_2}^2 + M_{h_3}^2) - (M_{h_2}^2 - M_{h_3}^2) \cos 2 \beta}
{4 v_3^{2}}\,, \nn \\
\lambda_{23} &=& \frac{(M_{h_2}^2 - M_{h_3}^2) \sin \beta \cos\beta}{v_2 v_3}\,, \nn \\
\mu_2^2 &=& \lambda_2 v_2^2 + \lambda_{23} \frac{v_3^2}{2}\,, \nn \\
\mu_3^2 &=& \lambda_3 v_3^2 + \lambda_{23} \frac{v_2^2}{2}\,.
\end{eqnarray}

\subsection{Extended Neutrino Sector and Inverse seesaw}
\label{iss}
Here we have shown only those terms in the Lagrangian
for the neutrino sector which are necessary for the inverse
seesaw mechanism. All the terms in the Lagrangian are
allowed by both $\ubl$ and $\umt$ gauge symmetries. 
\begin{eqnarray}
\mathcal{L}_{N} &\supset & 
\dfrac{i}{2}\sum_{\alpha = e,\,\mu,\,\tau}
\overline{N_{\alpha}} \gamma^{\mu}D^{N}_{\mu} N_{\alpha} 
+ \overline{N_{\alpha}^{\prime}} \gamma^{\mu}D^{N^{\prime}}_{\mu} N_{\alpha}^{\prime}
+ \overline{\nu_{\alpha}}\gamma^{\mu}D^{\nu}_{\mu} \nu_{\alpha}
-\left(\sum_{\alpha=e,\,\mu,\,\tau}\dfrac{M_{\alpha\alpha}}{2}
\overline{N_{\alpha}^{c}}N_{\alpha}^{\prime}\right. \nn \\ && \left. 
+ h_{e \mu} \overline{N_{e}^{c}}N_{\mu}^{\prime}\phi_3 
+ h_{e \tau}\overline{N_{e}^{c}}N_{\tau}^{\prime}\phi_3^{\dagger}
+ h_{\mu e}\overline{N_{\mu}^{c}}N_{e}^{\prime}\phi_3^{\dagger}  
+ h_{\tau e} \overline{N_{\tau}^{c}}N_{e}^{\prime}\phi_3 
+ y_{ee} \overline{N^{\prime\,c}_{e}} N_{e}^{\prime} \phi_2^{\dagger}\right.\nn \\ && \left.
+ y_{\mu\tau}\overline{N^{\prime\,c}_{\mu}}N_{\tau}^{\prime}\phi_2^{\dagger}
+ \sum_{\alpha=e,\,\mu,\,\tau} y_{\alpha}\, \overline{{L_{\alpha}}}\,
\tilde {\phi_{h}} N_{\alpha} +h.c.\right)\,,
\label{rh-lag}
\end{eqnarray}
where $D^{X}_{\mu}$ represents the covariant derivative for the field
$X$ ($X=N_\alpha,\,N^{\prime}_{\alpha},\,\nu_{\alpha}$). The first three
terms are the kinetic terms for $N_\alpha,\,N^{\prime}_{\alpha}$
and $\nu_{\alpha}$, while the last term is the Yukawa interaction
term (Dirac type) between the SM lepton doublet
($L_{\alpha} = \left(\nu_{\alpha}~~ l_{\alpha}\right)^{T}$),
Higgs doublet and the RH neutrino
$N_{\alpha}$. All the other terms in the above Lagrangian are the
interaction terms between $N_{\alpha}$, $N^{\prime}_{\beta}$
and the Majorana mass terms of $N^{\prime}$. 
The general form of the inverse seesaw Lagrangian is given by,  
\begin{eqnarray}
L_{ISS} &=& \sum_{\alpha, \beta = e, \mu, \tau}
m_{D}^{\alpha \beta} \overline{\nu}_{\alpha}
N_{\beta}  + \overline{N_{\alpha}^{c}} M_{N}^{\alpha \beta}
N_{\beta}^{\prime} + \overline{N_{\alpha}^{\prime \,c}}
\mu^{\alpha \beta} N_{\beta}^{\prime} + h.c.
\label{iss-lag}
\end{eqnarray}
Therefore, from the above Lagrangian one can construct a $9 \times 9$ mass
matrix sandwiched between the basis states
$\left(\overline{\nu_{\alpha}}~~\overline{N^c_{\alpha}}~~
\overline{{N^{\prime}}^c_{\alpha}} \right)$ and
$\left(\nu^c_{\beta}~~N_{\beta}~~N_{\beta}^{\prime}\right)^T$ as  
\begin{eqnarray}
\mathcal{M} = \left(\begin{array}{ccc}
0 ~~&~~ m_{D}
~~&~~0 \\
~~&~~\\
m_{D}^{T} ~~&~~ 0
~~&~~ M_{N} \\
~~&~~\\
0 ~~&
~~ M_{N}^{T} ~~&~~ \mu \\
\end{array}\right) \,.
\label{mncomplex1}
\end{eqnarray}
Comparing Eq.~(\ref{iss-lag}) with the Lagrangian $\mathcal{L}_{N}$,
one can easily find the structure of the individual
matrices, namely, $m_D$, $M_N$ 
and $\mu$ as
\begin{eqnarray}
m_{D} = \left(\begin{array}{ccc}
\frac{y_e}{\sqrt{2}}v ~~&~~ 0
~~&~~0 \\
~~&~~\\
0 ~~&~~ \frac{y_\mu}{\sqrt{2}}v
~~&~~ 0 \\
~~&~~\\
0 ~~&
~~ 0 ~~&~~ \frac{y_\tau}{\sqrt{2}}v \\
\end{array}\right) \,,
\label{dirac-mass}
\end{eqnarray}

\begin{eqnarray}
{M}_{N} = \left(\begin{array}{ccc}
M_{ee} ~~&~~ \dfrac{ \vmt}{\sqrt{2}} h_{e \mu}
~~&~~\dfrac{\vmt}{\sqrt{2}} h_{e \tau} \\
~~&~~\\
\dfrac{\vmt}{\sqrt{2}} h_{\mu e} ~~&~~ M_{\mu\mu}^{R} + i\,M_{\mu\mu}^{I}
~~&~~ 0 \\
~~&~~\\
\dfrac{\vmt}{\sqrt{2}} h_{\tau e} ~~&
~~ 0 ~~&~~ M_{\tau\tau}^{R} + i M_{\tau\tau}^{I} \\
\end{array}\right) \,,
\label{majorana-complex-mass}
\end{eqnarray}

\begin{eqnarray}
\mu = \left(\begin{array}{ccc}
\frac{y_{ee}}{\sqrt{2}}v_2 ~~&~~ 0
~~&~~0 \\
~~&~~\\
0 ~~&~~ 0
~~&~~ \frac{y_{\mu\tau}^R + i\,y_{\mu\tau}^I}{\sqrt{2}}v_2 \\
~~&~~\\
0 ~~&
~~ \frac{y_{\mu\tau}^R + i\,y_{\mu\tau}^I}{\sqrt{2}}v_2 ~~&~~ 0 \\
\end{array}\right) \,.
\label{mu-mat-complex-mass}
\end{eqnarray}
Although all the elements allowed by the imposed gauge symmetries
can be in general complex numbers, however by redefining the phases
of the fermionic fields one can check that there can only be
three independent complex phases possible. Consequently, we have chosen
(2,2), (3,3) elements of $M_N$ matrix and (2,3)\footnote{(3,2) element
of $\mu$ is also a complex number due to the symmetric nature of
Majorana mass matrix $\mu$.} element of $\mu$
matrix as complex numbers.

In the above mass matrix given by Eq.~(\ref{mncomplex1}), for simplicity
we have neglected the Majorana mass term $M_R$ of $N_{\alpha}$
in the (2,2) element of
$\mathcal{M}$, although
it is allowed by both $\ubl$ and $\umt$ symmetries. This is because,
if we consider $M_R$ with the same order of magnitude as $M_N$,
then in the limit $M_{N} > m_{D} >> \mu $ (the condition for
inverse seesaw mechanism), this term 
 has a negligible effect on the light neutrino
mass matrix \cite{Zhou:2012ds}.
 
After diagonalising the $9 \times 9$ mass matrix $\mathcal{M}$
we get the three light neutrinos and six heavy neutrinos with
the following expressions of mass matrices, 
\begin{eqnarray}
m_{\nu_l} & = & m_{D} M_{N}^{-1} \mu (M_{N}^{T})^{-1} m_{D}^{T}\,,\nn \\
m^2_{N_{H}} , m^2_{N_{H^{\prime}}} & = & m^2_{D} + M^2_{N}  \,.
\end{eqnarray}
The physical basis ($\nu^c_{l}$, $N_{H}$, $N_{H^{\prime}}$)
can be written in terms of the ($\nu^c$, $N$, $N^{\prime}$)
basis in the following manner \cite{Khalil:2010iu},
\begin{eqnarray}
\nu_{l}^c & = & \nu^c + a_1 N + a_2 N^{\prime}, \nn \\
N_{H} & = & a_3 \nu^c + \kappa N - \kappa N^{\prime}, \nn \\
N_{H^{\prime}} & = & \kappa N + \kappa N^{\prime}\,,
\end{eqnarray}
where $a_{1,2} \sim m_{D}/(M_{N} \sqrt{2 + \frac{2m_{D}}{M_{N}}})$,
$a_3 \sim m_{D}/M_{N}$ and $\kappa \sim \sin(\pi/4)$.
In determining $a_1$, $a_2$ and $a_3$ we have taken the model parameters
value as given in the Appendix \ref{App:AppendixA}.
 
Here $m_{D}$, $M_{N}$ and $\mu$ have a very particular structure due to the
($\mu-\tau$) flavour symmetry.
For our convenience, we have defined few new variables which are,
\begin{eqnarray}
Y_{\alpha} = \frac{y_{\alpha}}{\sqrt{2}} v,\,\, 
V_{\alpha \beta} = \frac{v_{\mu \tau}}{\sqrt{2}} h_{\alpha \beta}\,\, {\rm and}\,\,
Y_{\alpha \beta} = \frac{y_{\alpha \beta}}{\sqrt{2}} v_2\,,
\end{eqnarray}
where, all the parameters defined above have dimensions of mass.
In section \ref{neutrino-res}, we will show
the allowed regions among the different parameters of the above mentioned
mass matrices (Eq.\,(\ref{dirac-mass}) - (\ref{mu-mat-complex-mass}))
after applying the neutrino oscillation data constraints for both NH and IH.
Constraints on mixing angles, mass square
differences and the sum of all the light neutrinos, 
which we have followed in determining the allowed
parameter space are as follows,
\begin{itemize}
\item there is a bound on the sum of all three light neutrinos from cosmology which is,
$\sum_i m_{i} < 0.23$ eV at $2\sigma$ C.L. \cite{Ade:2015xua},
\item mass squared differences for NH (IH) are $6.93\, (6.93) <\dfrac{\Delta m^2_{21}}
{10^{-5}}\,{\text{eV}^2} < 7.97\, (7.97)$ and $2.37\, (2.33) <\dfrac{\Delta m^2_{31(13)}}
{10^{-3}}\,{\text{eV}^2} < 2.63\, (2.60)$ in $3\sigma$ range \cite{Capozzi:2016rtj},
\item all three mixing angles for NH (IH) are $30^{\circ}\, (30^{\circ}) <\,
\theta_{12}\,<36.51^{\circ}\, (36.51^{\circ})$,
$37.99^{\circ}\, (38.23^{\circ}) <\,\theta_{23}\,< 51.71^{\circ}\, (52.95^{\circ})$ and
$7.82^{\circ}\, (7.84^{\circ})<\,\theta_{13}\,<9.02^{\circ}\, (9.06^{\circ})$
also in $3\sigma$ range
\cite{Capozzi:2016rtj}.
\end{itemize}   
In the above mass matrices $m_{D}$, $\mathcal{M}_{R}$ and $\mu$, many elements are
zero. Therefore, when we will apply the above constraints, the oscillation data
will put severe constraints on the parameter values and we will get nice
correlations among the parameters which we will see in the result section.
\subsection{Dark Matter Sector}
\label{dm-lag}
As discussed earlier, we need to introduce four chiral fermions
($\xi_L$, $\eta_L$, $\chi_{1R}$ and $\chi_{2R}$) with fractional
${\rm B-L}$ and ${\rm L_{\mu}-L_{\tau}}$ charges (see Table \ref{tab1})
to make the present model anomaly free. The Lagrangian for these
exotic fermionic states has been denoted by
$\mathcal{L}_{DM}$ in Eq.~(\ref{Lblmt}). Since these chiral fermions
are singlet under the SM gauge group, hence $\mathcal{L}_{DM}$
contains only those terms which are invariant under $\ubl\times\umt$
gauge group: 
\begin{eqnarray}
\mathcal{L}_{DM} &=& i\Bigg[\,\overline{\xi_{L}}\,\gamma^{\mu}\left(\partial_{\mu}
+i\,\frac{4}{3} \gbl {\zbl}_{\mu} +i\,\frac{1}{3} \gmt {\zmt}_{\mu} \right) \xi_{L} 
+\,\overline{\eta_{L}}\,\gamma^{\mu}\left(\partial_{\mu} + i \frac{1}{3}
\gbl\,{\zbl}_{\mu} + i \frac{4}{3}
\gmt\,{\zmt}_{\mu} \right) \eta_{L} \nn \\ && 
+\overline{\chi_{1R}}\gamma^{\mu}
\left(\partial_{\mu} - i\frac{2}{3}\gbl{\zbl}_{\mu} 
+ i\frac{1}{3}\gmt{\zmt}_{\mu} \right) \chi_{1R}
+\overline{\chi_{2R}}\gamma^{\mu}
\left(\partial_{\mu} - i\frac{2}{3}\gbl{\zbl}_{\mu} 
+ i\frac{4}{3}\gmt{\zmt}_{\mu} \right) \chi_{2R} \Bigg]
\nn \\ &&
- \left(\gamma_{1}\,\overline{\xi_{L}} \chi_{1\,R}\,\phi_2 + 
\kappa_{1}\,\overline{\eta_{L}} \chi_{2\,R}\,\phi_1 + h.c.\right)\,,
\label{Ldm}
\end{eqnarray}
The last term in the above equation shows that 
in order to write Dirac mass terms for all the chiral fermions,
one needs to have at least two scalar fields $\phi_1$ and $\phi_2$
with different ${\rm B-L}$ charges. This is the main reason
why we have introduce more than one scalar field to break the
$\ubl$ symmetry spontaneously.
After symmetry breaking of $\ubl$ symmetry by the VEVs of $\phi_1$
and $\phi_2$, the Dirac mass matrix with respect to the basis
states ($\xi_L$,~$\eta_L$) and (${\chi_1}_R$,~${\chi_2}_R$)
takes the following diagonal form, which is possible
due to the assignment of particular ${\rm L_{\mu}-L_{\tau}}$
charges to these chiral fermions.
 \begin{eqnarray}
\mathcal{L}^{DM}_{mass} = 
( \overline{\xi_{L}} & \overline{\eta_{L}})
\left(\begin{array}{cc}
\gamma_1 \frac{v_2}{\sqrt{2}} ~~&~~ 0 \\
0 ~~&~~ \kappa_1 \frac{v_1}{\sqrt{2}}\\
\end{array}\right)
\left(\begin{array}{c}
\chi_{1\,R} \\ \chi_{2\,R}
\end{array}\right)\,.
\label{dm-mass}
\end{eqnarray} 
From the above mass matrix one can easily construct
two Dirac-type fermionic states $\Sigma_1 = \xi_L \bigoplus \chi_{1\,R}$
and $\Sigma_2=\eta_L \bigoplus \chi_{2\,R}$
which are the combinations of these exotic chiral fermions
and having masses $M_1=\gamma_1 \frac{v_2}{\sqrt{2}}$
and $M_2 = \kappa_1 \frac{v_1}{\sqrt{2}}$, respectively.
Between these two Dirac fermions, the lightest one will be
stable and can be a dark matter candidate.
Therefore, for definiteness, throughout this
work we have considered $M_1<M_2$ and hence
$\Sigma_1$ is our dark matter candidate.
Moreover, we can also write down
the Yukawa couplings $\gamma_1$ and $\kappa_1$ in terms
of masses $M_1$, $M_2$ of the fermions $\Sigma_1$, $\Sigma_2$
in the following way,
\begin{eqnarray}
\gamma_1 = \frac{\sqrt{2} M_1}{v_2} \,,\,\,\, 
\kappa_1 = \frac{\sqrt{2} M_2}{v_1}\,.
\label{higgs-coupling}
\end{eqnarray} 
\subsubsection{\bf Spin Independent scattering cross section of $\Sigma_1$}
\begin{figure}[h!]
\centering
\includegraphics[angle=0,height=5cm,width=7cm]{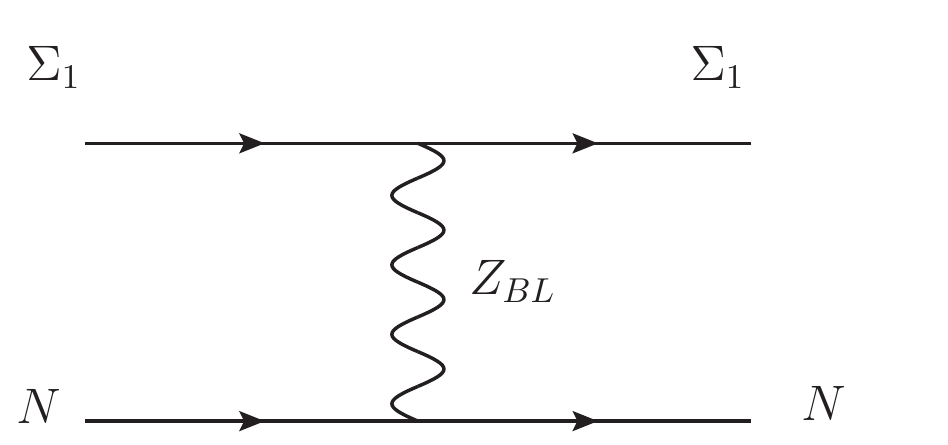}
\caption{Feynman diagram for the spin independent elastic scattering
cross section between dark matter candidate $\Sigma_1$ and the nucleon
($N$) mediated by $\zbl$.}
\label{si-dd-diagram}
\end{figure}

In this work, our dark matter candidate $\Sigma_1$ can talk to the quark
sector only through the exchange of ${\rm B-L}$ gauge boson $\zbl$, as we
have not considered the mixing between the SM Higgs boson and the
other BSM singlet scalars. The coupling between $\Sigma_1$ and $\zbl$ has
the following form,
\begin{eqnarray}
g_{\overline{\Sigma_1}\,\Sigma_1\zbl} = \gamma^{\mu}
\,\left(a_{\overline{\Sigma_1}\,\Sigma_1\zbl} + 
b_{\overline{\Sigma_1}\,\Sigma_1\zbl} \gamma_5\right) \,,
\end{eqnarray}
where,
\begin{eqnarray}
a_{\overline{\Sigma_1}\,\Sigma_1\zbl} = -\dfrac{\gbl}{3}\,,~~~
b_{\overline{\Sigma_1}\,\Sigma_1\zbl} = {\gbl}\,.
\end{eqnarray}
As shown in Fig.\,\ref{si-dd-diagram},
our dark matter candidate scatters off the detector nucleus via
a $t$-channel process mediated by $\zbl$. The
expression for the spin independent elastic scattering
cross section for the above process is given by,
\begin{eqnarray}
\sigma_{SI} = \frac{\mu^{2}}{\pi}\,
\frac{g^2_{\overline{N}N\zbl} a^2_{\overline{\Sigma_1}\,\Sigma_1\zbl}}
{M^4_{Z_{BL}}}\,,
\end{eqnarray}
where $\mu$ is the reduced mass of nucleon ($N$) and dark matter
$\Sigma_1$ given as $\mu = \frac{M_{DM} M_{N}}{M_{DM} + M_{N}}$.
The quantity $g_{\overline{N}N\zbl}$ is the effective coupling
between $N$ and $\zbl$, which is defined as
$g_{\overline{N}N\zbl}\,\overline{N}\gamma^{\mu}N\,{\zbl}_{\mu}$
and it has the following expression,
\begin{eqnarray}
g_{\overline{N}N\zbl} &=&
\sum_{q = u,d} f^N_{V_q}\,g_{\overline{q}q\zbl}\,.
\end{eqnarray}
Here, $g_{\overline{q}q\zbl} = \dfrac{\gbl}{3}$
represents the coupling between first generation quark and
the gauge boson $\zbl$. Now, $f^N_{V_u} = 2$, $f^N_{V_d}=1$
for $N=p$ (proton) and
$f^N_{V_u} = 1$, $f^N_{V_d}=2$ for $N=n$ (neutron) \cite{Belanger:2008sj}.
Therefore, the effective coupling between nucleon ($p$ or $n$)
and $\zbl$ is
\begin{eqnarray}
g_{\overline{N}N\zbl} &=& 3 \times \dfrac{\gbl}{3}\,.
\end{eqnarray}
In Section \ref{Sec:DM}, we show the variation of
$\sigma_{SI}$ with the mass of dark matter and we 
compare our results with the latest bounds on $\sigma_{SI}$
from XENON1T \cite{Aprile:2017iyp} and PandaX-II \cite{Cui:2017nnn}
dark matter direct search experiments.
\section{results}
\label{result}
\label{muon_g-2}
\subsection{Muon ($g-2$)}
\begin{figure}[h!]
\centering
\includegraphics[angle=0,height=6cm,width=8cm]{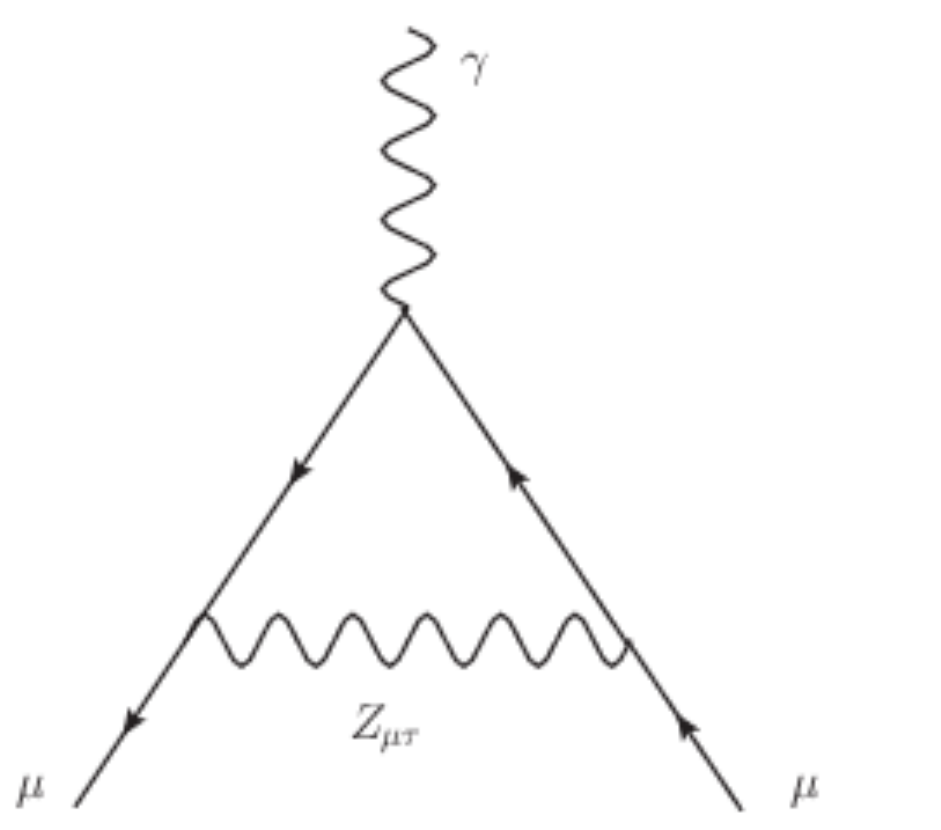}
\caption{One loop Feynman diagram for muon ${\rm (g-2)}$
contribution mediated by extra gauge boson $Z_{\mu \tau}$. }
\label{muong2}
\end{figure}
The presence of the extra neutral gauge boson $Z_{\mu\tau}$ gives an additional
one-loop contribution to the muon ($g-2$) shown in Fig.\,\ref{muong2}. 
The contribution coming from the digram in Fig.\,\ref{muong2} to the
muon magnetic moment $a_{\mu}^{the}$ is given as,
\begin{eqnarray}
\Delta a_{\mu}(Z_{\mu \tau}) = \dfrac{g_{\mu \tau}^{2}}{8 \pi^{2}}
\int_{0}^{1} dx \dfrac{2 x(1-x)^{2}}{(1-x)^{2} + rx},
\label{intg2}
\end{eqnarray}
where, $r = (M_{Z_{\mu \tau}}/m_{\mu})^{2}$ and $M_{Z_{\mu \tau}}$ is the mass of $Z_{\mu \tau}$.
Further, $m_{\mu}$ is the mass of muon $\mu^{\pm}$ while $g_{\mu \tau}$ is the
U(1)$_{\rm L_{\mu} - L_{\tau}}$ gauge coupling.
For $g_{\mu \tau} = 9 \times 10^{-4}$ and $M_{Z_{\mu \tau}} = 0.1$ GeV,
one can get from Eq.~(\ref{intg2}) 
\begin{eqnarray}
\Delta a_{\mu} = 2.257 \times 10^{-9} \,,
\end{eqnarray} 
which lies roughly within the 3.2\,$\sigma$ range of the observed discrepancy. 
In what follows, we keep $g_{\mu \tau}$ and $M_{Z_{\mu \tau}} $ fixed at the above values.
Here, we would like to note that the benchmark value we have chosen,
is allowed from the neutrino trident production experiments like CHARM-II and CCFR
\cite{Geiregat:1990gz, Mishra:1991bv, Altmannshofer:2014pba}.
As a result, following Eq.\,\,(\ref{massZmt}), we fix 
$v_{\mu\tau} = 111.11$ GeV throughout the analysis.
We will see that these parameter values, as we have also argued before, affect
other phenomenology such as neutrino masses and mixing angles and DM. 
\subsection{Neutrino Parameters}
\label{neutrino-res}
\begin{figure}[h!]
\centering
\includegraphics[angle=0,height=7.5cm,width=8.0cm]{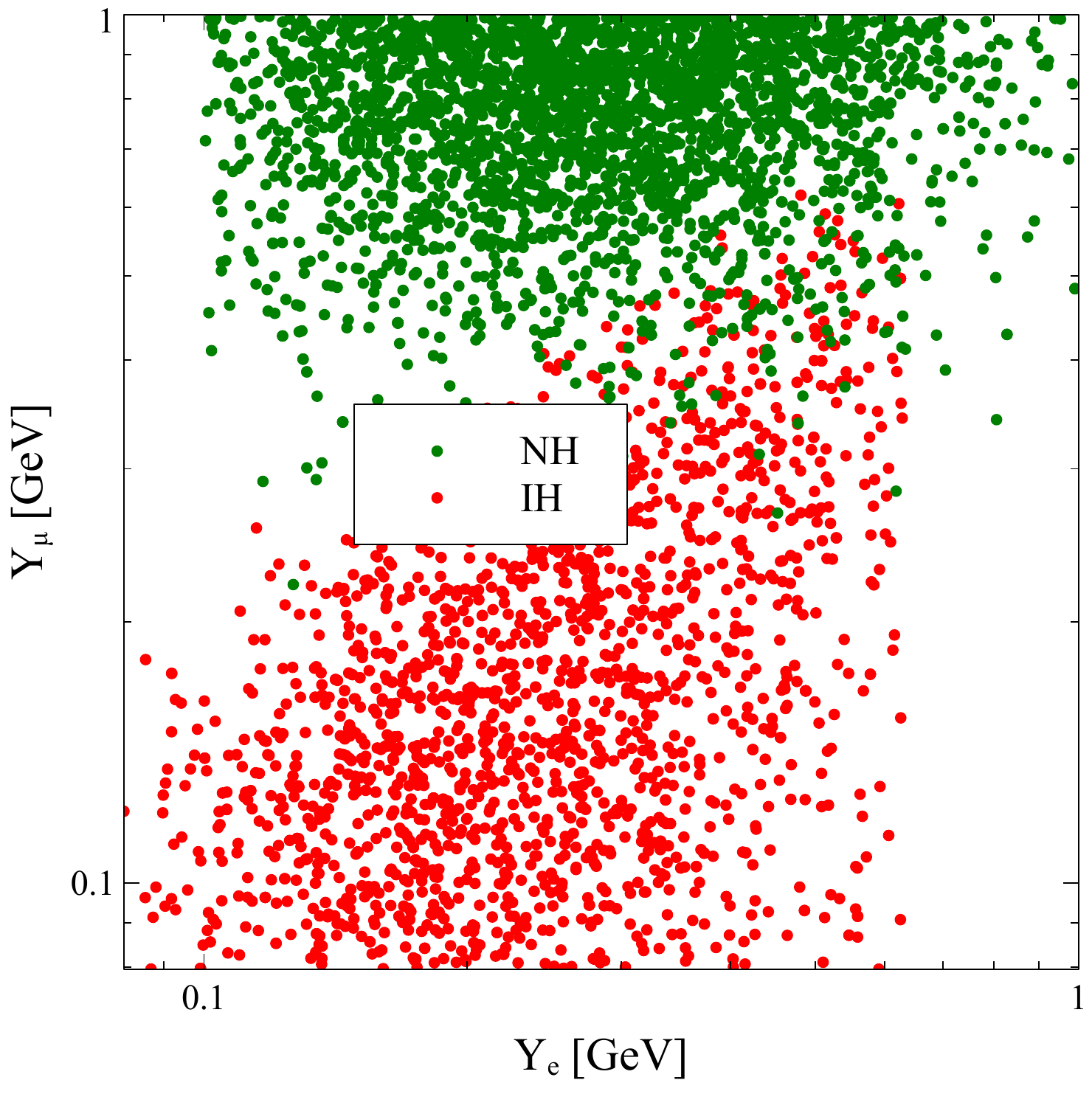}
\includegraphics[angle=0,height=7.50cm,width=8.0cm]{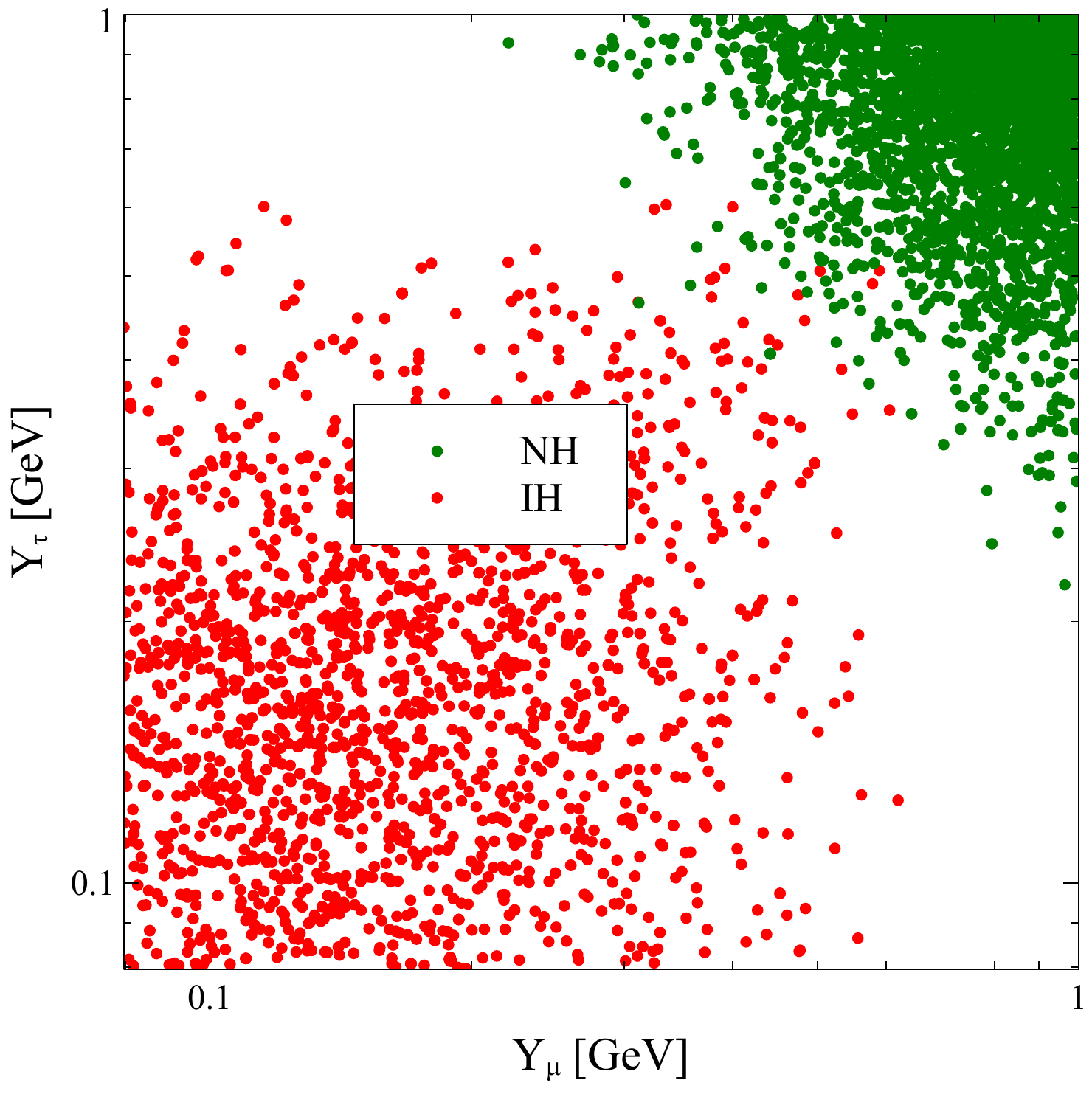}
\includegraphics[angle=0,height=7.5cm,width=8.0cm]{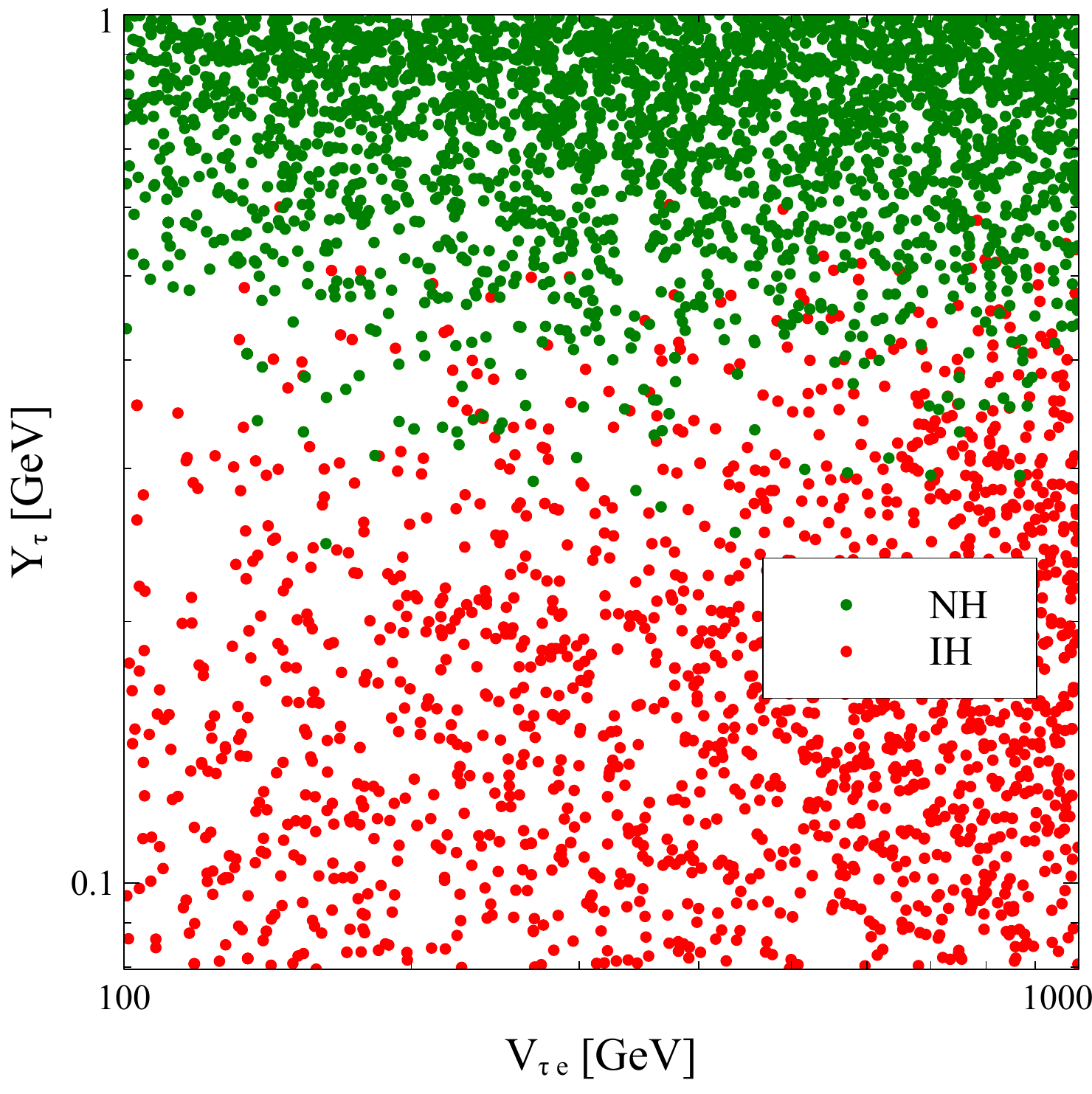}
\includegraphics[angle=0,height=7.50cm,width=8.0cm]{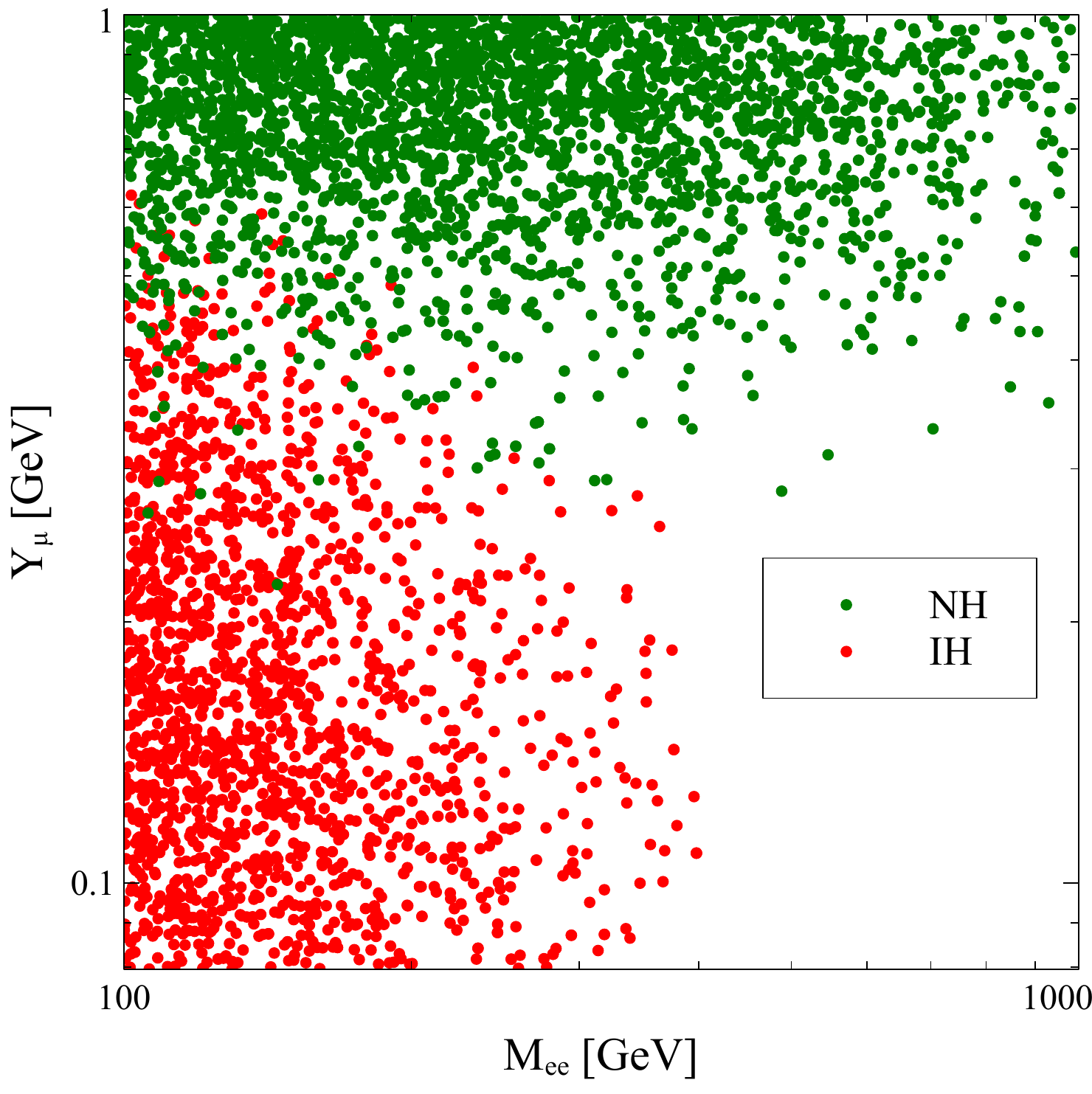}
\caption{Each point satisfies neutrino oscillation data
and there is no overlap between the allowed parameter
spaces for NH and IH. All the parameters have been
scanned over the ranges as displayed
in Eq.\,(\ref{parameter-range})}.      
\label{scatt-1}
\end{figure}

In generating the scatter plots among the different parameters
of the neutrino mass matrices (given in the
Eqs.\,(\ref{dirac-mass}-\ref{mu-mat-complex-mass}),
we have varied the model parameters in the following
range both for NH and IH: 
\begin{eqnarray}
0.1\,\, {\rm [GeV]}\,\,& < \,\,\, m_{D}(i,j)\,\,\,
<&\,\, 1\,\, {\rm [GeV]}\,,\nn \\
100\,\,{\rm [GeV]}\,\, &<\,\,\, \mathcal{M}_{R}(i,j)\,\,\,
<&\,\, 1000\,\,{\rm [GeV]}\,, \nn \\
10^{-7}\,\, {\rm [GeV]}\,\,& < \,\,\, m_{D}(i,j)\,\,\,
<&\,\, 10^{-5}\,\, {\rm [GeV]}\,.
\label{parameter-range}
\end{eqnarray}
After satisfying the neutrino oscillation data as mentioned in the
section \ref{iss}, we get the allowed regions for the model
parameters which are described below.

In Fig.\,\ref{scatt-1}, we have shown the variation of the parameters
in the planes $Y_{\mu}-Y_{e}$, $Y_{\tau}-Y_{\mu}$, $Y_{\tau}-V_{\tau e}$
and $Y_{\mu}-M_{ee}$. We can see that the neutrino oscillation data 
puts constraints on the model parameters, restricting them to take values 
in the ranges shown in Fig.\,\ref{scatt-1}. More interestingly, we note that 
in all the plots shown in this figure, the neutrino oscillation 
data prefers regions of parameter space that are almost completely distinct for the 
NH and IH cases. 
 
\begin{figure}[]
\centering
\includegraphics[angle=0,height=7.5cm,width=8.0cm]{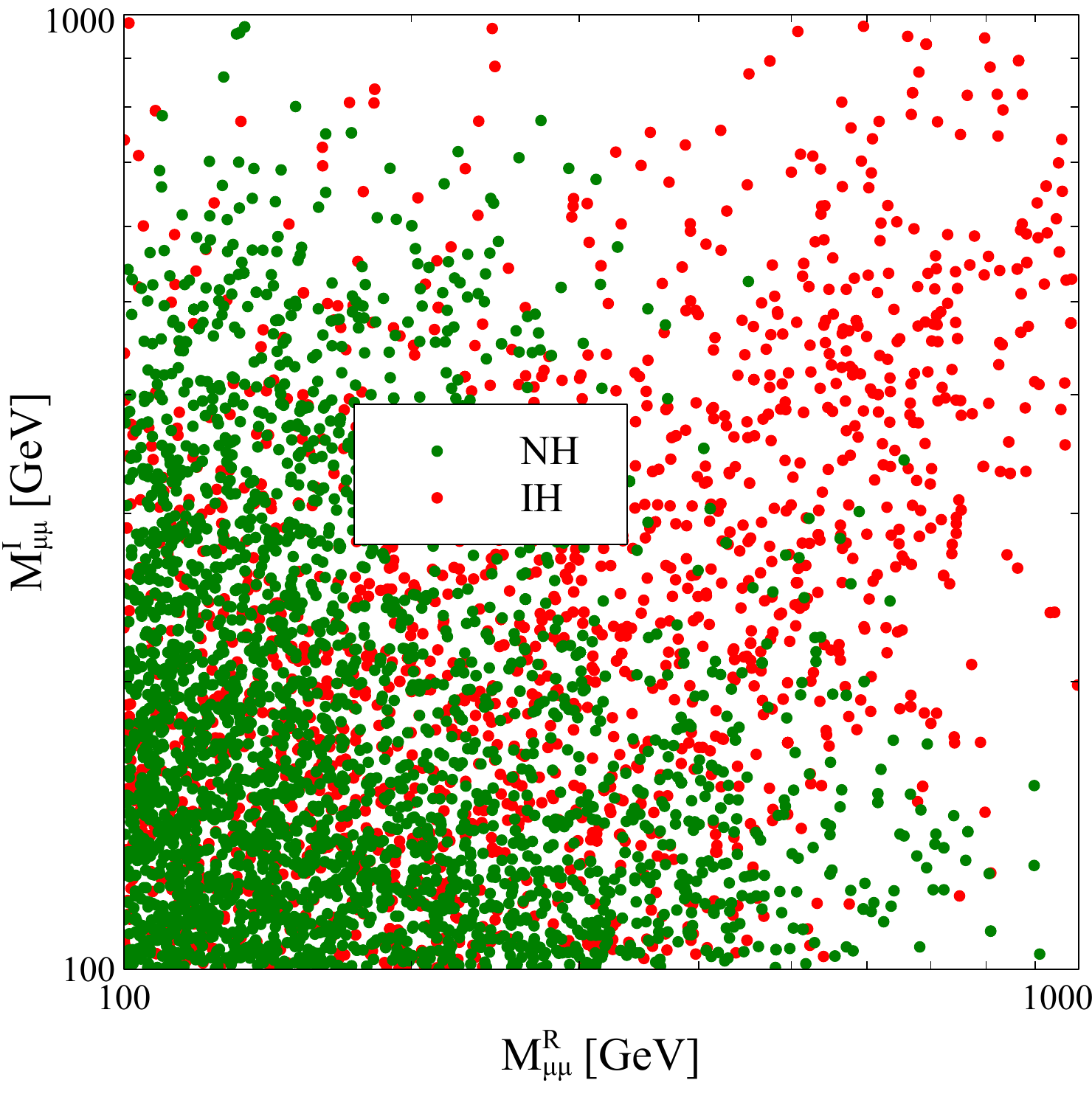}
\includegraphics[angle=0,height=7.50cm,width=8.0cm]{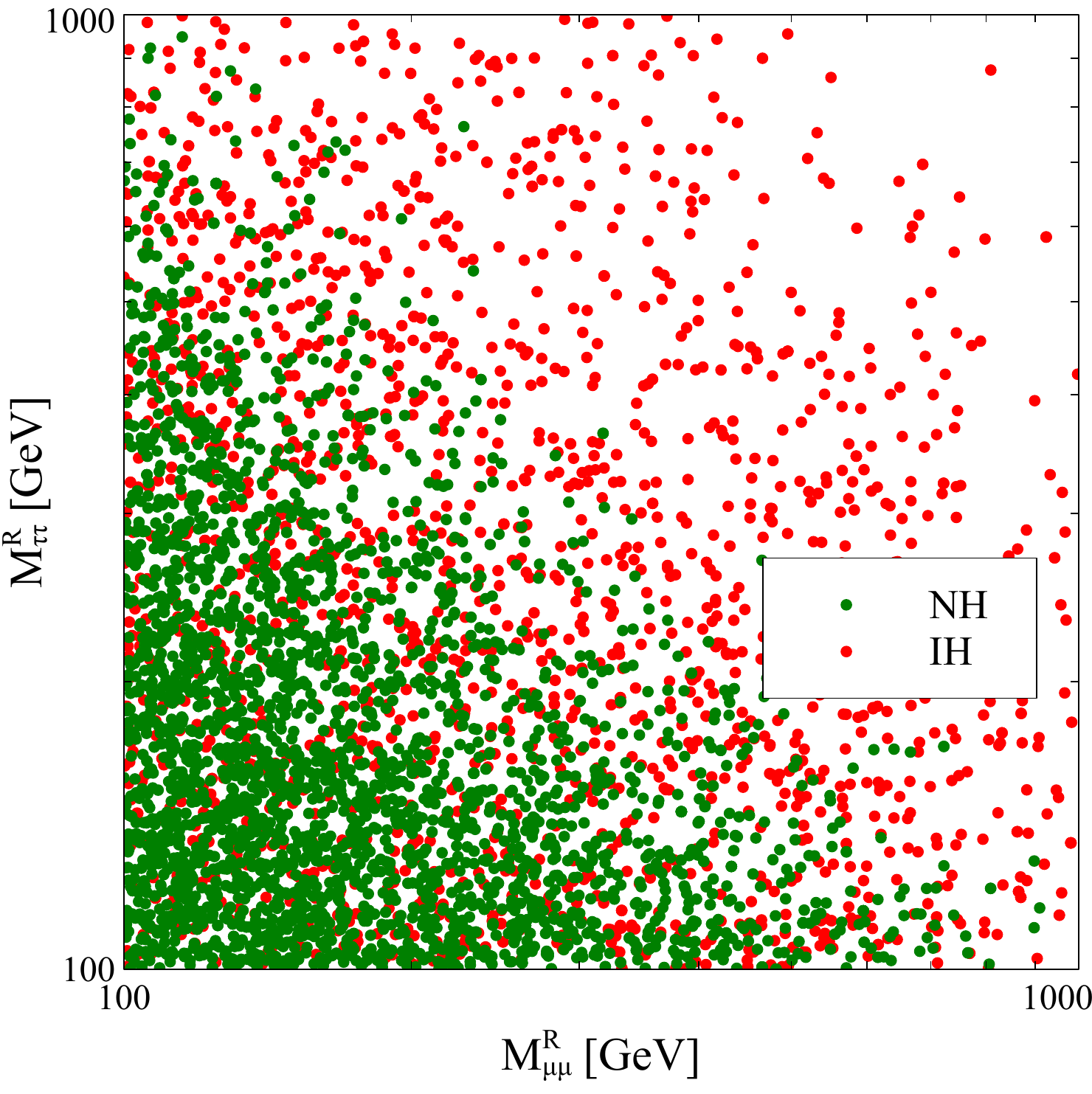}
\includegraphics[angle=0,height=7.5cm,width=8.0cm]{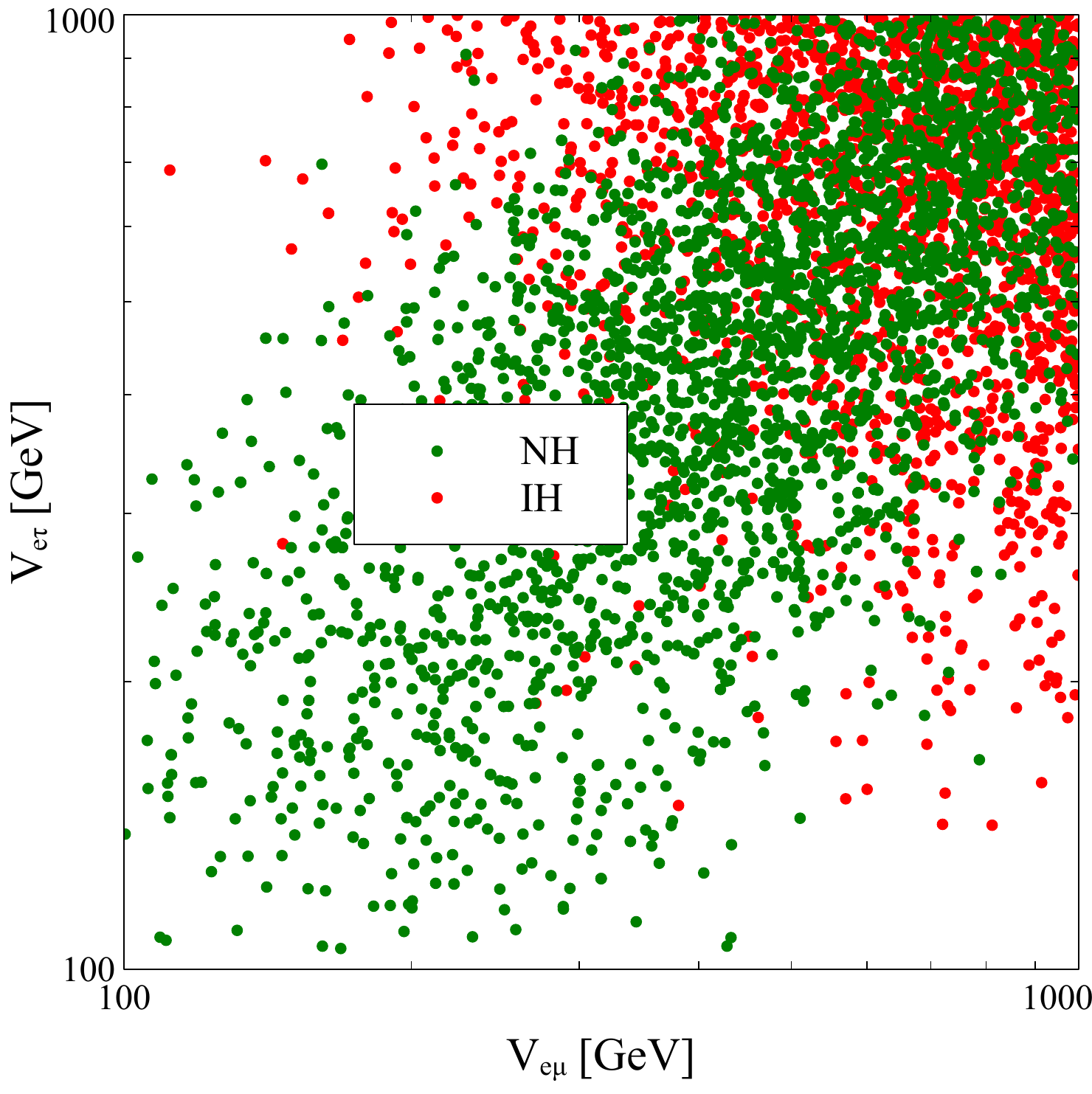}
\includegraphics[angle=0,height=7.50cm,width=8.0cm]{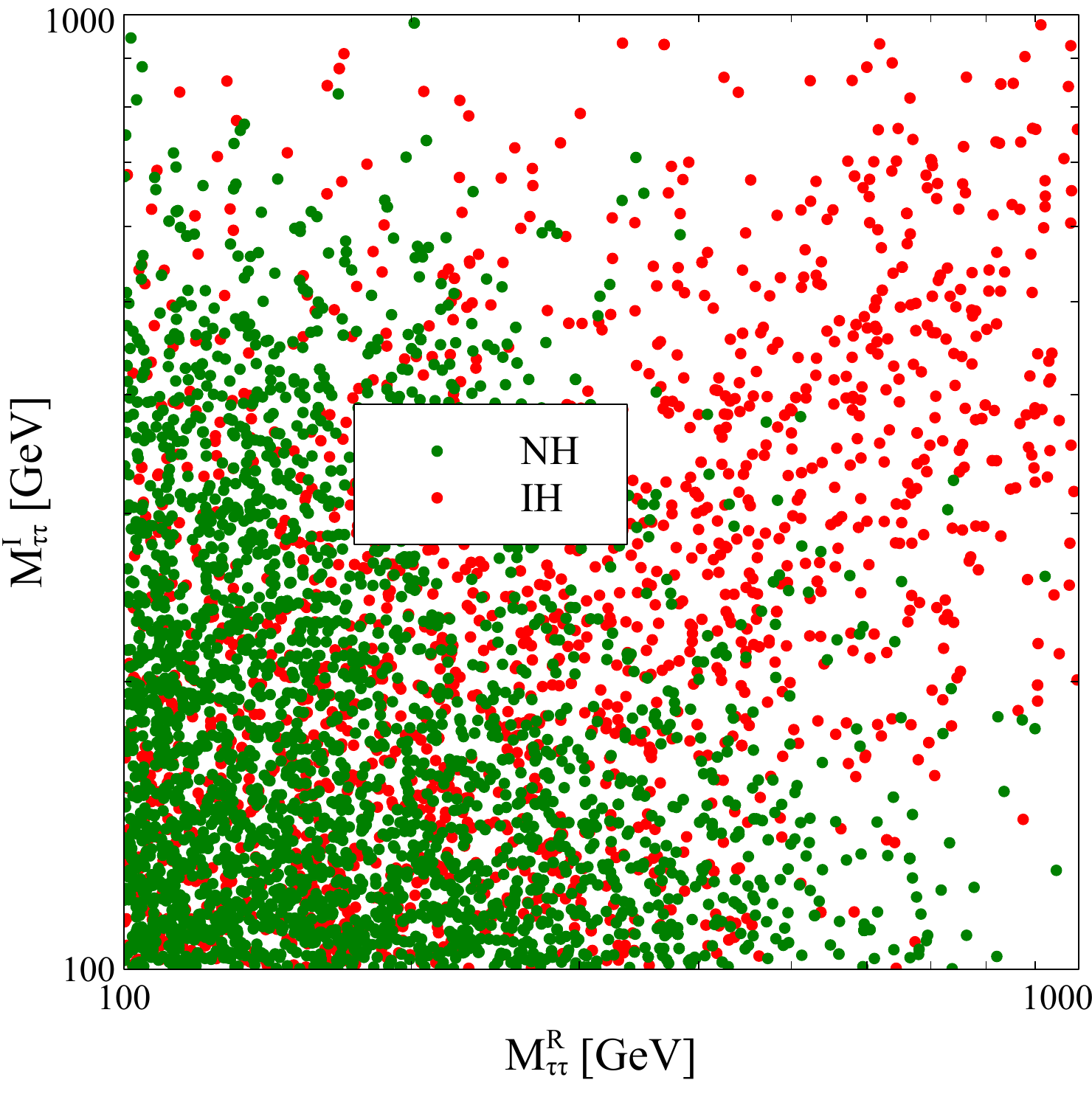}
\caption{Each point satisfies neutrino oscillation data 
and both the allowed regions for NH and IH are overlapping
on each other. Similar to Fig.\,\,\ref{scatt-1}, here also all
the parameters have been scanned over the
entire ranges given in Eq.\,(\ref{parameter-range})}      
\label{scatt-2}
\end{figure}

In Fig.\,\ref{scatt-2} we shown scatter plots in the 
$M^I_{\mu\mu}-M^R_{\mu\mu}$, $M^R_{\mu\mu}-M^R_{\tau\tau}$,
$V_{e\mu}-V_{e\tau}$ and $M^R_{\tau\tau}-M^I_{\tau\tau}$ planes. 
Contrary to Fig.\,\ref{scatt-1}, in Fig.\,\ref{scatt-2}
the allowed parameter values for NH and IH are seen to overlap.
As before, here too we get nice correlation between the parameter 
values on putting the observational constraints from the neutrino 
oscillation data. 
One interesting point to note here is that the planes $M^R_{\mu\mu}-M^I_{\mu\mu}$
and $M^R_{\tau\tau}-M^I_{\tau\tau}$ behave in a similar way
for both NH and IH.

\begin{figure}[]
\centering
\includegraphics[angle=0,height=7.5cm,width=8.0cm]{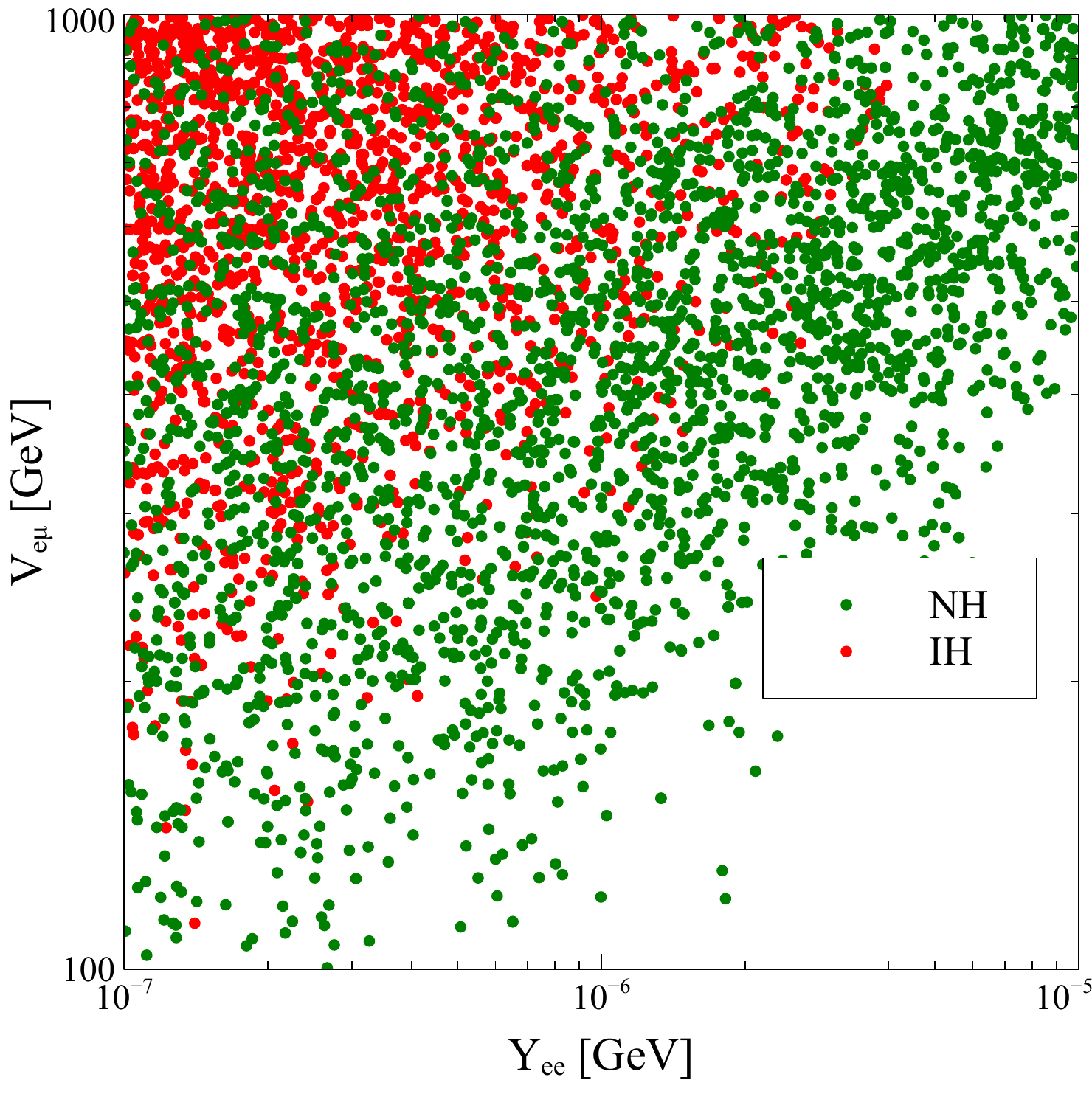}
\includegraphics[angle=0,height=7.50cm,width=8.0cm]{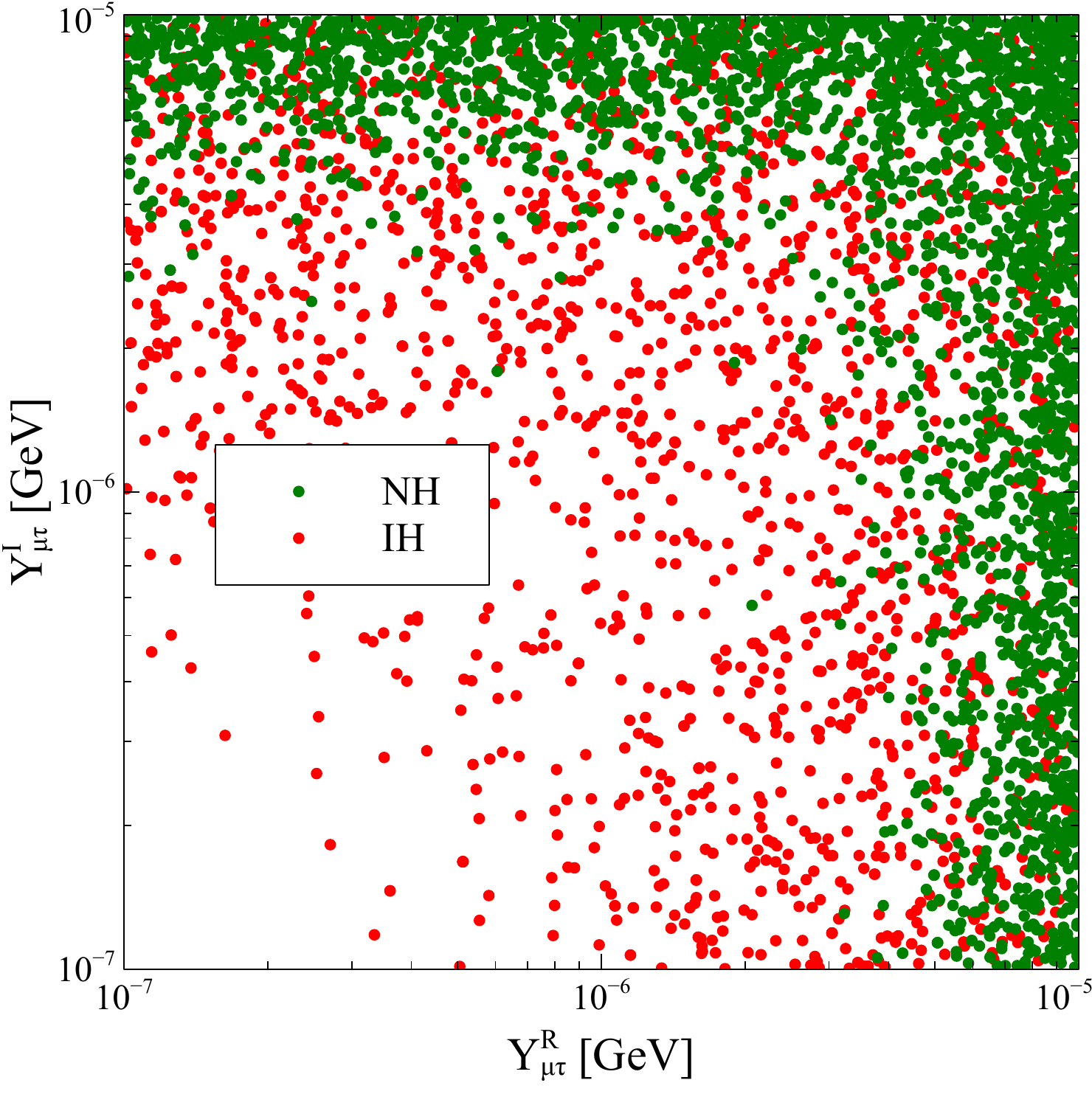}
\caption{Each point satisfies neutrino oscillation data
and here also we find overlapping regions between NH and IH
for the elements of $\mu$ matrix.}   
\label{scatt-3}
\end{figure}

In  Fig.\,\ref{scatt-3}, we show the allowed regions in the 
$Y_{ee} - V_{e\mu}$ and $Y^R_{\mu \tau} - Y^I_{\mu\tau}$ model space 
for NH and IH. The correlation between the parameters for both NH and IH can be seen.

From the above study we can infer that the elements of the Dirac mass matrix
$m_{D}$ play an important role in determining the neutrino mass hierarchy
for the light neutrinos. One can
notice in the planes $Y_e - Y_{\mu}$ and $Y_{\mu} - Y_{\tau}$
of Fig.\,\ref{scatt-1}, that these parameters have different values for
NH and IH. For other parameters of the $\mathcal{M}_{R}$
and $\mu$ matrices, there exist overlap regions
in different planes between the elements of this two matrices. Therefore, 
these parameters are less important in determining the neutrino mass hierarchy, 
unlike the Dirac mass matrix elements as discussed above. 

\subsection{Dark Matter}
\label{Sec:DM}

\begin{figure}[h!]
\centering
\includegraphics[angle=0,height=8cm,width=11cm]{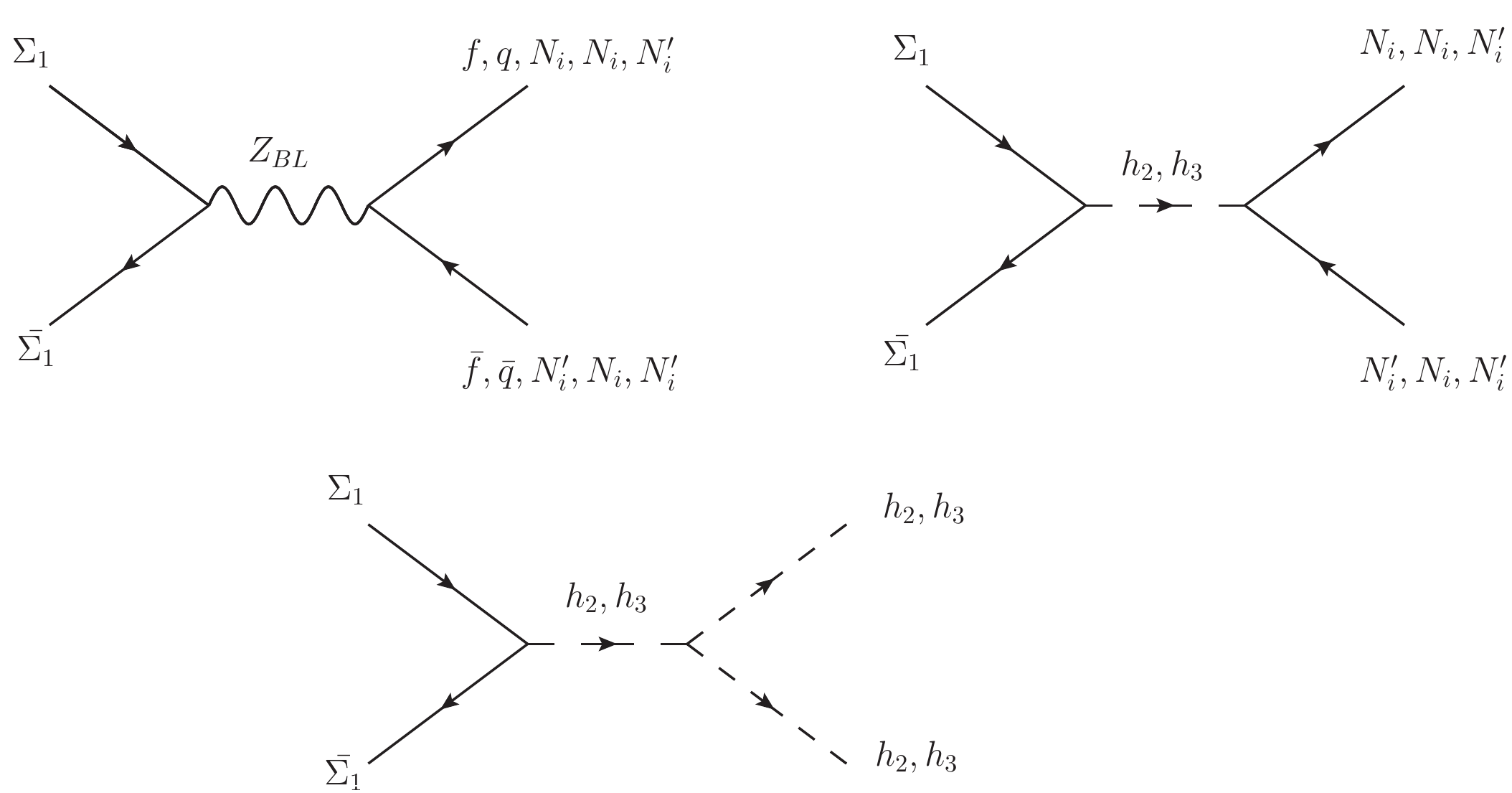}
\caption{Feynman diagrams for the annihilation of dark matter.}
\label{dm-feyn}
\end{figure}

As discussed in the subsection \ref{dm-lag},
among the two neutral fermions $\Sigma_1$ and $\Sigma_2$,
the lightest one will be the 
DM candidate. In our analysis we have considered $\Sigma_1$
as the DM candidate with mass $M_{DM} (=M_1)$. The most
important property of any DM candidate, which has been
measured precisely by WMAP and Planck, is its relic density.
The DM relic density $\Omega {h^2}$, which is defined as the ratio of
DM mass density to the critical density of the Universe, is
related to the DM comoving number density at the present epoch
by the following relation \cite{Edsjo:1997bg},
\begin{eqnarray}
\Omega h^2 = 2.755\times 10^8 \left(\dfrac{M_{DM}}{\rm GeV}
\right) Y(T_0)\,,
\end{eqnarray}
where $Y(T_0)$ is the value of comoving number density
$Y$ at $T=T_0$, the present temperature of the Universe. 
In order to find $Y(T_0)$, one has to solve the relevant
Boltzmann equation which is given by \cite{Gondolo:1990dk}
\begin{eqnarray}
\frac{dY}{dx} =
-\left(\frac{45\,G_{N}}{\pi}\right)^{-\frac{1}{2}}
\frac{M_{DM}\,\sqrt{g_\star}}{x^2}\,
\frac{1}{2} \langle{\sigma {\rm{v}}}\rangle
\left(Y^2-(Y^{eq})^2\right)\,.
\label{be1}
\end{eqnarray}
Here, $x=\frac{M_{DM}}{T}$ and $\langle\sigma {\rm v} \rangle$ is the thermally
averaged annihilation cross section of DM into various
final state particles. Moreover, $G_{N}$ is the Newton's
gravitational constant. Further, the quantity $g_{\star}$ is
related to the degrees of freedom $g_{eff}$ and $h_{eff}$
of energy and entropy densities of the Universe and
its expression is given in Ref. \cite{Gondolo:1990dk}.
The 1/2 factor in the R.H.S. of Boltzmann equation
is due to non-self-conjugate nature of our DM candidate
$\Sigma_1$ (Dirac fermion). 
We have solved the Boltzmann equation numerically using
\texttt{micrOMEGAs} \cite{Belanger:2013oya} package. For that,
we have generated the required model files by implementing
present model in \texttt{Feynrules} \cite{Alloul:2013bka}.
In Fig.\,\ref{dm-feyn}, we show the dominant annihilation
channels of $\Sigma_1$, which are mediated by $h_2$, $h_3$
and $Z_{BL}$ respectively.

\begin{figure}[h!]
\centering
\includegraphics[angle=0,height=7.5cm,width=8.0cm]{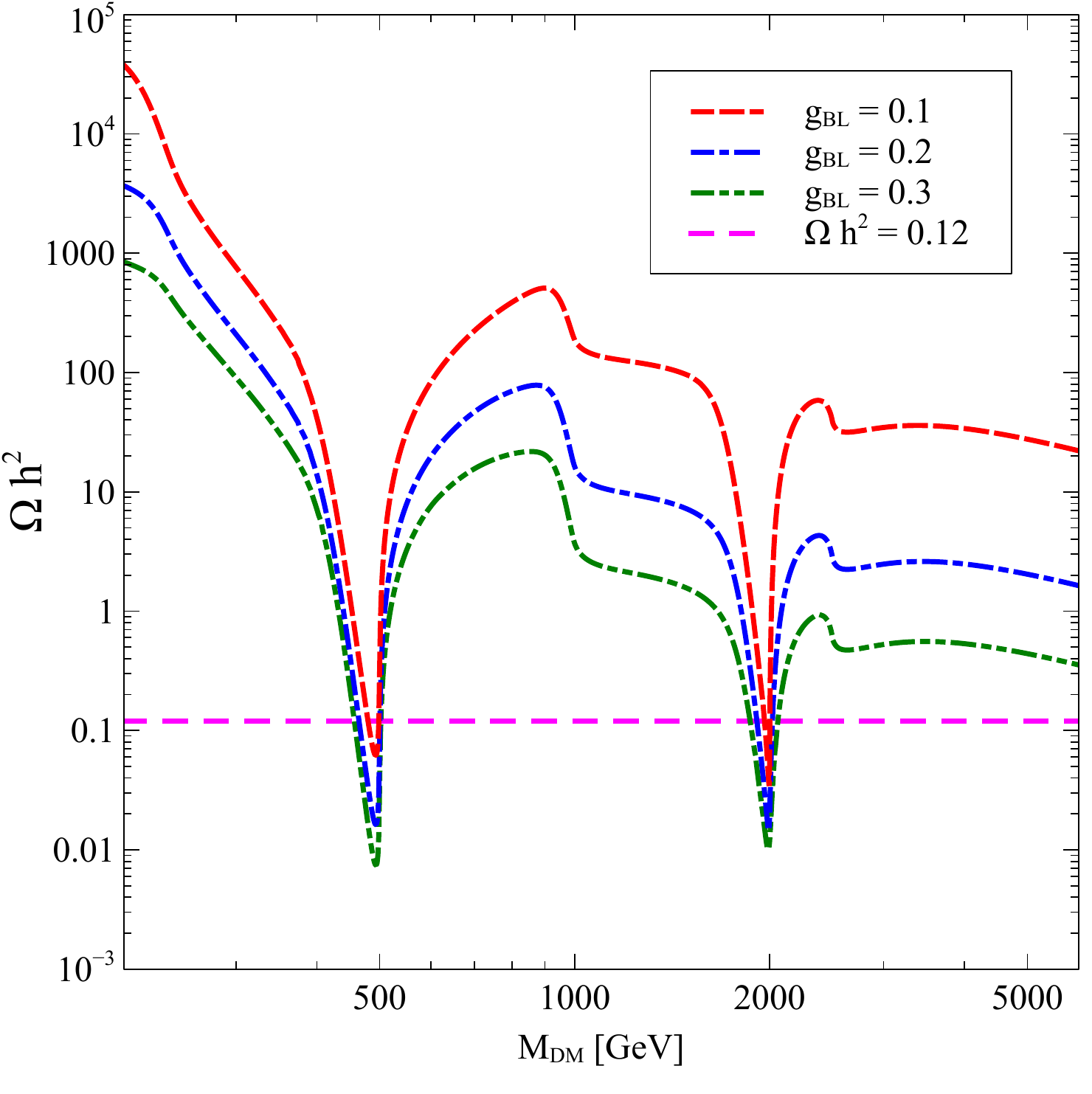}
\includegraphics[angle=0,height=7.50cm,width=8.0cm]{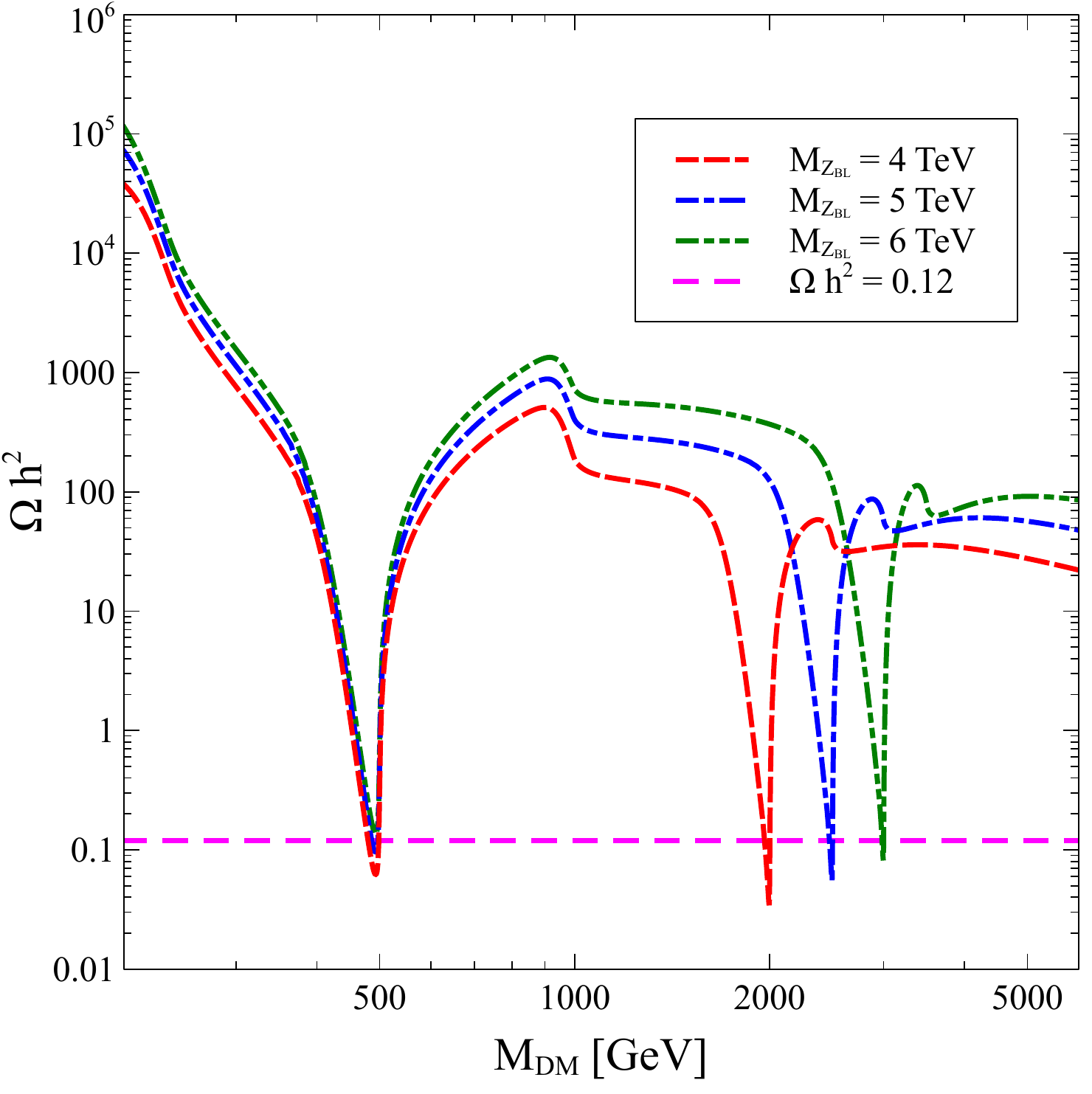}
\caption{Left (Right) panel: Variation of relic density with the DM mass
for three different values of gauge coupling (gauge boson mass).
While other BSM parameters have been kept fixed at $r_{{\it vev}} = 5$,
$M_{N_1} = M_{N^{\prime}_1} = 361.4$ GeV,
$M_{N_2} = M_{N^{\prime}_2} = 296.2$ GeV,
$M_{N_3} = M_{N^{\prime}_3} = 111.0$ GeV, $M_{h_2} = 1000.0$ GeV, 
$M_{h_3} = 250$ GeV, $M_{Z_{BL}} (g_{BL}) = 4000$ (0.1) GeV,
$h_{e\mu} = 1.58$, $h_{e\tau} = 1.63$, $h_{\mu e} = 2.38$, $h_{\tau e} = 1.34$,
$\sin \beta = 0.07$,
$\Delta M = M_2 - M_1 = M_2-M_{DM} = 50$ GeV.}      
\label{line-plot-1}
\end{figure}

In the left panel of Fig.\,\ref{line-plot-1}, we show the variation of DM relic
density with the mass of dark matter $M_{DM}$ for three different values of
U(1)$_{B-L}$ gauge coupling. The figure shows two
resonance regions corresponding to the $h_2$ and $Z_{BL}$ masses, respectively.
The dependence of the DM relic density on $g_{BL}$ is similar in both the 
resonance regions.
For the $Z_{BL}$ resonance region, the cross-section increases with the increase of $g_{BL}$
and as a result the relic density decreases, as can be seen 
in the figure. On the other hand, for the resonance region corresponding to the 
$h_2$ mediated diagrams, the effect of 
$g_{BL}$ comes indirectly. We see from Eq.\,(\ref{v2}) that with 
increase of $g_{BL}$, $v_2$ decreases. Since the coupling $\gamma_1$ in Eq.\,(\ref{higgs-coupling})
depends on the  VEV $v_2$, the $\gamma_1$ coupling increases when $g_{BL}$
increases and consequently the cross-section of the $h_2$ mediated diagrams increase and the 
relic density falls. In the right panel of same figure
we have shown the variation of relic density for three
different values of gauge boson mass $M_{Z_{BL}}$, shown in the
legend. For different value of $M_{Z_{BL}}$
there is a shift of the resonance peak, as expected. 
Since the mass of $h_2$ is kept fixed, there is no visible change in the $h_2$ resonance peaks.
However, one can notice that around the $h_2$
resonance region the DM relic density increases almost linearly with the gauge boson mass.
Again, Eq.\,(\ref{v2}) shows that $v_2$ is 
proportional to gauge boson mass $M_{Z_{BL}}$, leading to this dependence, as shown by 
Eq.\,(\ref{higgs-coupling}).

\begin{figure}[h!]
\centering
\includegraphics[angle=0,height=7.5cm,width=8.0cm]{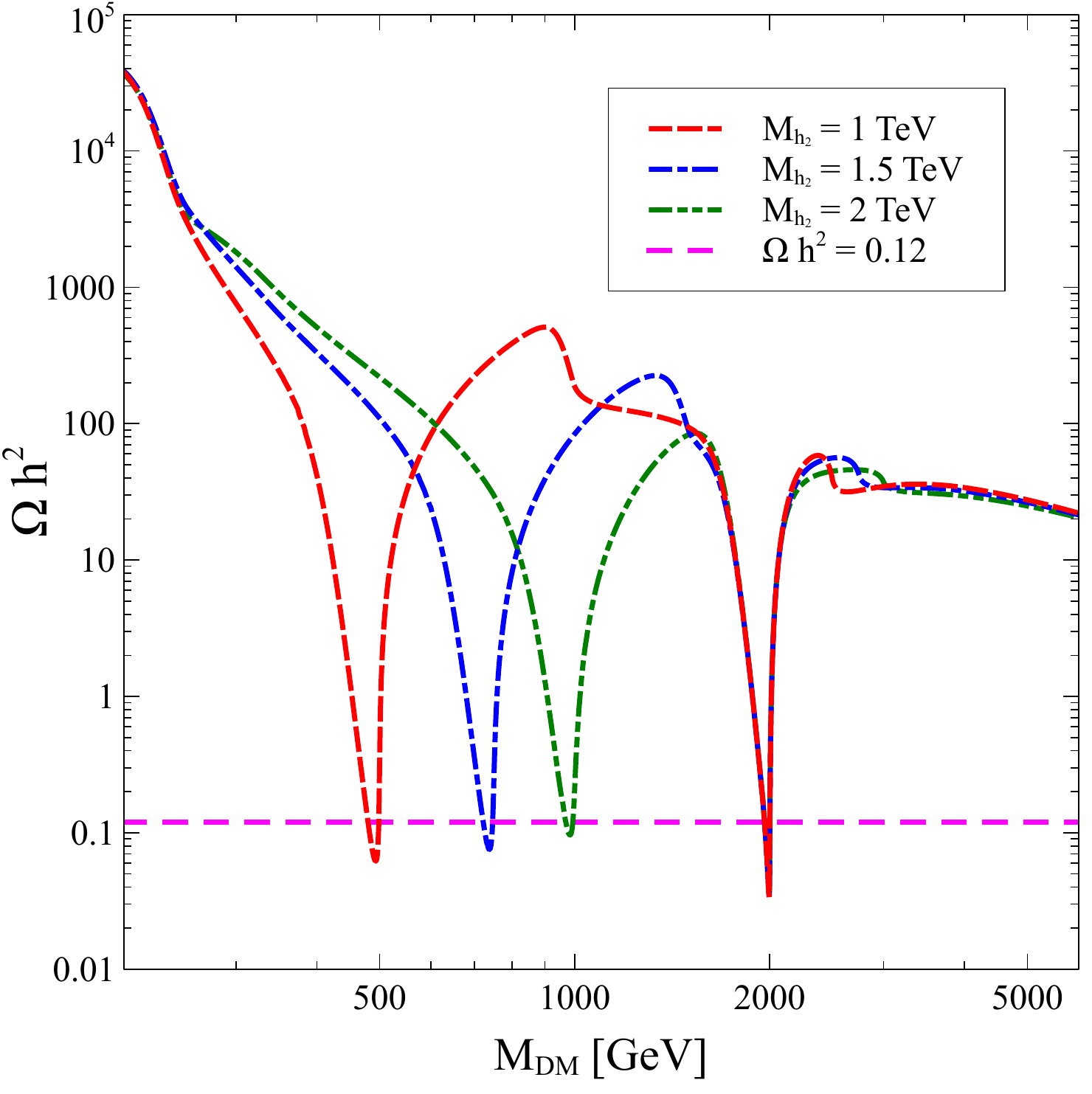}
\includegraphics[angle=0,height=7.50cm,width=8.0cm]{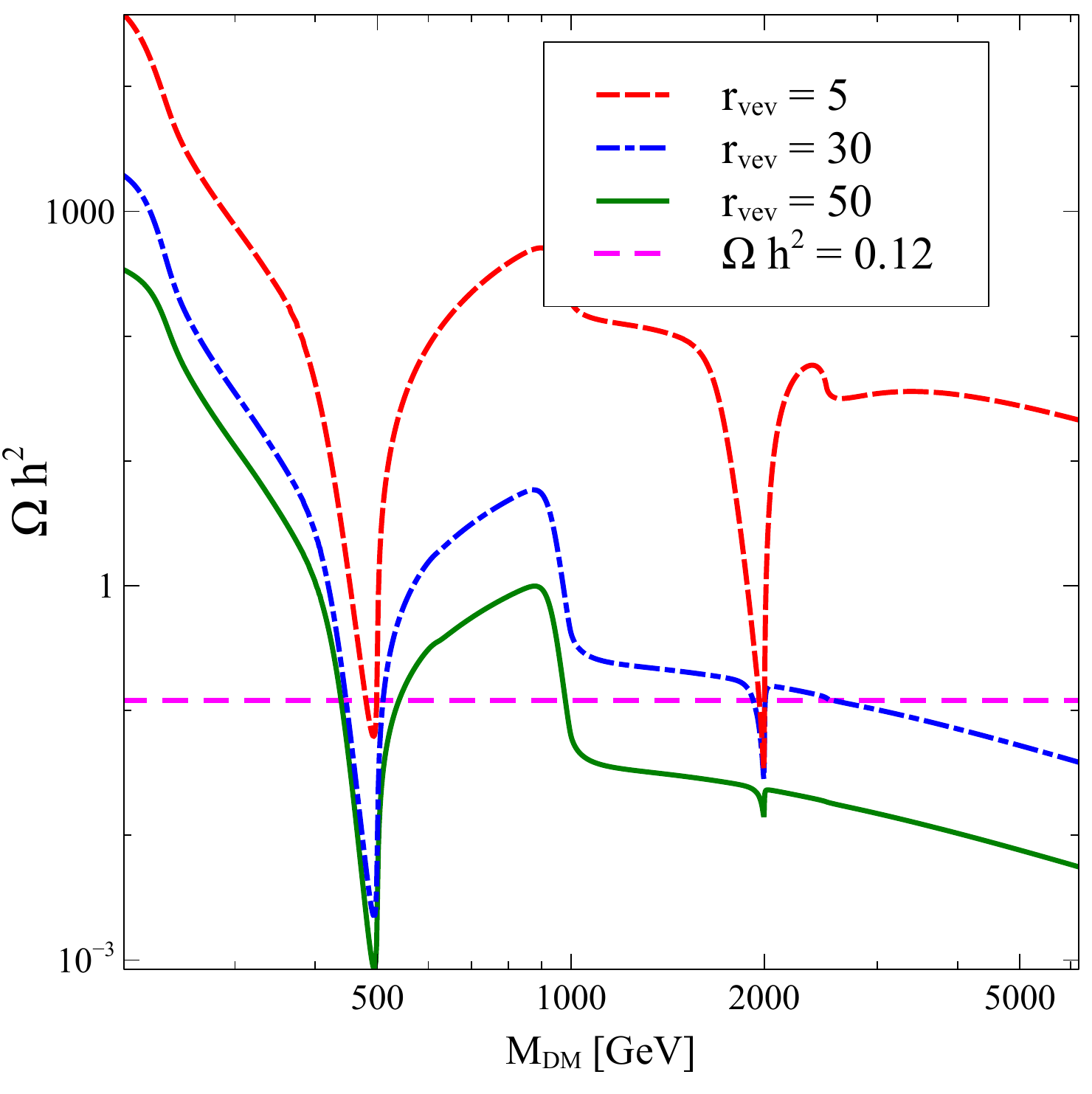}
\caption{Left (Right) panel: Variation of relic density with the DM mass
for three different values of $h_2$ mass (VEV ratio).
While other BSM parameters have been kept fixed at $r_{{\it vev}} = 5$,
$M_{N_1} = M_{N^{\prime}_1} = 361.4$ GeV,
$M_{N_2} = M_{N^{\prime}_2} = 296.2$ GeV,
$M_{N_3} = M_{N^{\prime}_3} = 111.0$ GeV, $M_{h_2} = 1000.0$ GeV, 
$M_{h_3} = 250$ GeV, $M_{Z_{BL}} (g_{BL}) = 4000$ (0.1) GeV,
$h_{e\mu} = 1.58$, $h_{e\tau} = 1.63$, $h_{\mu e} = 2.38$, $h_{\tau e} = 1.34$,
$\sin \beta = 0.07$,
$\Delta M = M_2 - M_1 = M_2-M_{DM} = 50$ GeV and for $a_1$, $a_2$
and $a_3$ see Appendix \ref{App:AppendixA}.}      
\label{line-plot-2}
\end{figure}

Similarly, in Fig.\,\ref{line-plot-2} we show the variation of
DM relic density with the DM mass. However, here in the left panel we show the variation
for three different values of $h_2$ mass. The $h_2$ resonance 
shift according to $M_{h_2}$. 
The $h_2$ mass does not have any impact on the $Z_{BL}$
resonance. On the other hand, in the
right panel we show the variation of DM relic density for three values of the 
VEV ratio. The figure shows that with the
increase of $r_{{\it vev}}$, the peak of the $Z_{BL}$ resonance
gradually disappears. The reason can be understood from Eq.\,(\ref{v2}).
With the increase of $r_{{\it vev}}$, for a fixed $v_1$, the value of $v_2$
decreases and consequently the $\gamma_1$ couplings increase.
This leads to greater dominance of the  $h_2$
mediated diagrams and as a result even in the $Z_{BL}$
resonance region it diminishes the gauge boson resonance effect.

\subsubsection{\bf Allowed parameter space near $h_2$ resonance}

\begin{figure}[h!]
\centering
\includegraphics[angle=0,height=7.5cm,width=8.0cm]{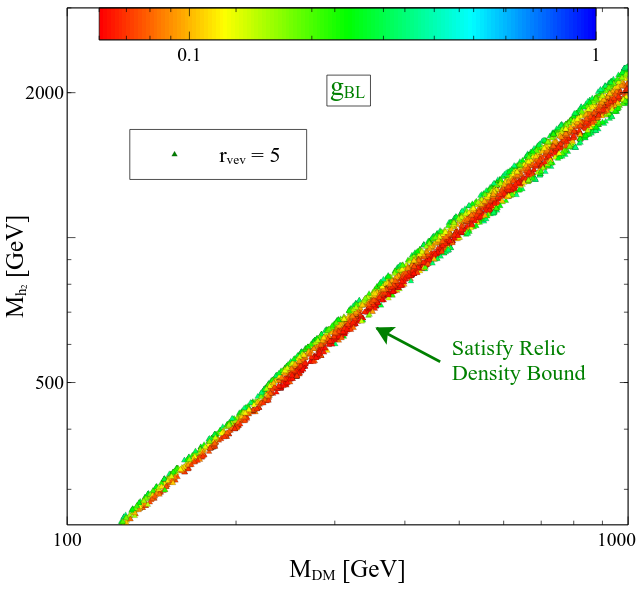}
\includegraphics[angle=0,height=7.50cm,width=8.0cm]{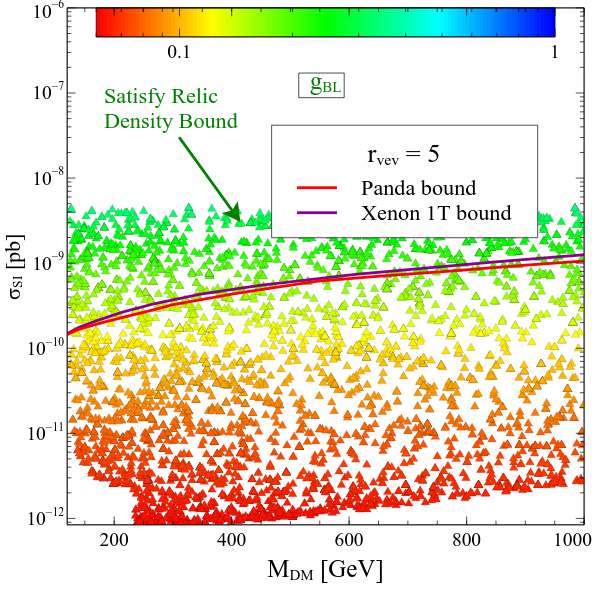}
\caption{Left (Right) Panel : Scatter plot in the $M_{DM} - M_{h_2}$
($M_{DM} - \sigma_{SI}$) plane after satisfying the DM relic density
bound. The parameters range have been shown in the Table \ref{tab-par-1} and other
parameters have been kept fixed at $r_{{\it vev}} = 5$,
$M_{N_1} = M_{N^{\prime}_1} = 361.4$ GeV,
$M_{N_2} = M_{N^{\prime}_2} = 296.2$ GeV,
$M_{N_3} = M_{N^{\prime}_3} = 111.0$ GeV, 
$M_{h_3} = 250$ GeV, $M_{Z_{BL}} = 4000$ GeV,
$h_{e\mu} = 1.58$, $h_{e\tau} = 1.63$, $h_{\mu e} = 2.38$, $h_{\tau e} = 1.34$,
$\sin \beta = 0.07$,
$\Delta M = M_2 - M_1 = M_2-M_{DM} = 50$ GeV.}       
\label{scatt-dm-1}
\end{figure}

\begin{table}[h!]
\begin{center}
\vskip 0.5cm
\begin{tabular} {||c||c||}
\hline
\hline
Parameters & Range\\
\hline
$M_{DM}$  & 150 - 1000 [GeV]\\
\hline
$M_{h_2}$  & 200 - 2500 [GeV]\\
\hline
$g_{BL}$ &
$10^{-3}$ - $1$\\
\hline
\hline
\end{tabular}
\end{center}
\caption{Parameters varied in generating the scatter plot near the $h_2$
resonance region.}
\label{tab-par-1}
\end{table}

In the left and right panels of Fig.\,\ref{scatt-dm-1}, 
the scatter plots in the $M_{DM} - M_{h_2}$ and $M_{DM} - \sigma_{SI}$
planes show the points that satisfy the DM relic density constraint.
In generating these plots we have varied three parameters as
shown in Table \ref{tab-par-1}.
We see in the left panel
a sharp correlation in $M_{DM} - M_{h_2}$ plane. This is expected
because the DM relic density is satisfied near the resonance region.
Another thing to note here is that for the lower value of 
$g_{BL}$ (can be seen from Fig.\,\ref{line-plot-1}), the relic density is satisfied 
in the narrower lower side of the resonance. In the left panel, lower values of
$g_{BL}$ are shown by the red points and one can see that the region
in $M_{DM} - M_{h_2}$ is narrower for these lower values of $g_{BL}$.
On the other hand, in the right panel we have shown the variation of
spin independent elastic scattering cross section with the
DM mass. In this work, DM can scatter elastically 
with the earth based detectors nuclei
only via the exchange of $Z_{BL}$.
Therefore, for the variation of $g_{BL}$,
the corresponding spin independent scattering
cross section cross section also changes.     
\subsubsection{{\bf Allowed parameter space near $Z_{BL}$ resonance}}

\begin{table}[h!]
\begin{center}
\vskip 0.5cm
\begin{tabular} {||c||c||}
\hline
\hline
Parameters & Range\\
\hline
$M_{DM}$  & 1000 - 3000 [GeV]\\
\hline
$M_{Z_{BL}}$  & 1800 - 6500 [GeV]\\
\hline
$g_{BL}$ &
$10^{-3}$ - $1$\\
\hline
$M_{DM}$ &
$< M_{Z_{BL}}$\\
\hline
\hline
\end{tabular}
\end{center}
\caption{Parameters varied in generating the scatter plot near the $Z_{BL}$
resonance region.}
\label{tab-par-2}
\end{table}

\begin{figure}[h!]
\centering
\includegraphics[angle=0,height=7.5cm,width=8.0cm]{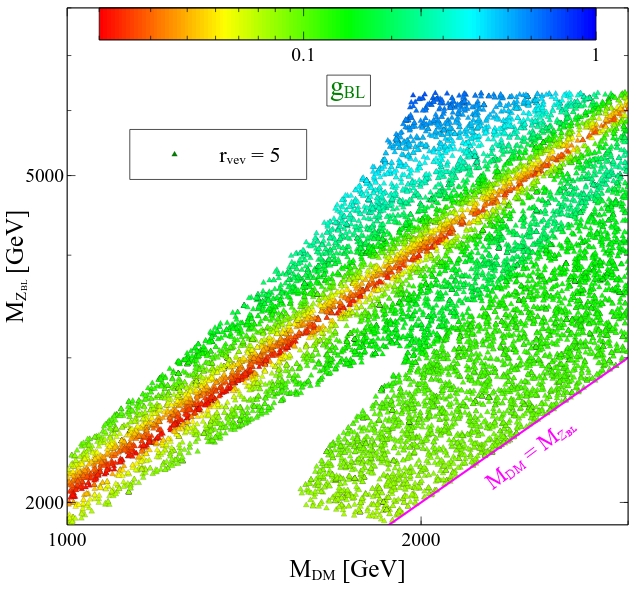}
\includegraphics[angle=0,height=7.50cm,width=8.0cm]{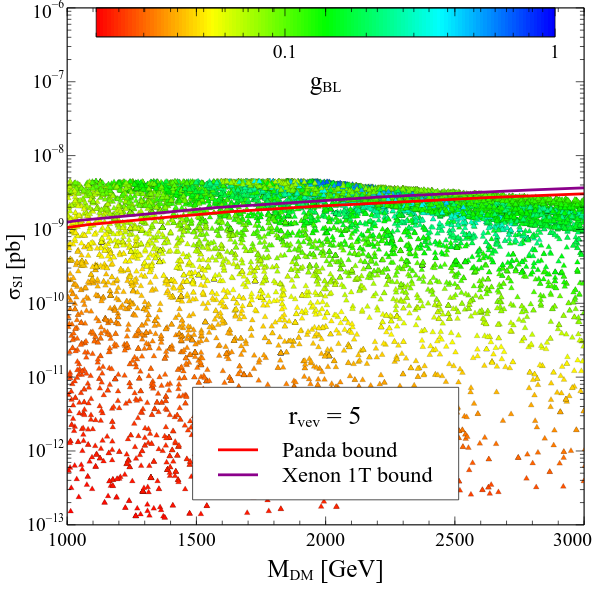}
\caption{Left (Right) Panel : Scatter plot in the $M_{DM} - M_{Z_{BL}}$
($M_{DM} - \sigma_{SI}$) plane after satisfying the DM relic density
bound. The parameters range have been shown in the Table \ref{tab-par-1} and other
parameters have been kept fixed at $r_{{\it vev}} = 5$,
$M_{N_1} = M_{N^{\prime}_1} = 361.4$ GeV,
$M_{N_2} = M_{N^{\prime}_2} = 296.2$ GeV,
$M_{N_3} = M_{N^{\prime}_3} = 111.0$ GeV, 
$M_{h_3} = 250$ GeV, $M_{h_2} = 1000$ GeV,
$h_{e\mu} = 1.58$, $h_{e\tau} = 1.63$, $h_{\mu e} = 2.38$, $h_{\tau e} = 1.34$,
$\sin \beta = 0.07$,
$\Delta M = M_2 - M_1 = M_2-M_{DM} = 50$ GeV.}      
\label{scatt-dm-2}
\end{figure}

In the left and right panel of Fig.\,\ref{scatt-dm-2}, we show the
points in the $M_{DM}-M_{Z_{BL}}$ and $M_{DM}-\sigma_{SI}$
planes that satisfy the DM relic density near the $Z_{BL}$
resonance region. In generating these plots three parameters have been
varied as shown in the Table \ref{tab-par-2}. In the left panel one can see that 
for the lower DM mass $M_{DM}$ we get a correlation between
$M_{DM}$ and $M_{Z_{BL}}$ upto DM mass of 1500 GeV. For higher
values of the DM mass $M_{DM}$, we get a broad region in the $M_{DM} - M_{Z_{BL}}$
plane. For the higher values of DM mass we see that the correlation breaks.
This is because for higher values of DM mass, relic density is satisfied for lower values of
$Z_{BL}$ mass as well. For lower values of $Z_{BL}$ mass,
the VEV $v_2$ takes small values (see Eq.\,(\ref{v2})), hence the $h_2$ mediated diagrams dominate and reduce
the effect of $Z_{BL}$ resonance region
(seen in the RP of Fig.\,\ref{line-plot-2}). 
In the right panel of the same figure we show the scatter plot
in the $\sigma_{SI}-M_{DM}$ plane that can be detected in the different
direct detection experiments
\cite{Akerib:2016vxi, Aprile:2015uzo, Aprile:2017iyp,
Cui:2017nnn, Aalbers:2016jon}.
In the same plane we have shown the
recent bound from XENON1T experiment \cite{Aprile:2017iyp} and
PandaX-II experiment \cite{Cui:2017nnn}.
A large area of the plane is accessible in the future
run of the different ongoing direct detection experiments.

\begin{figure}[h!]
\centering
\includegraphics[angle=0,height=7.50cm,width=8.0cm]{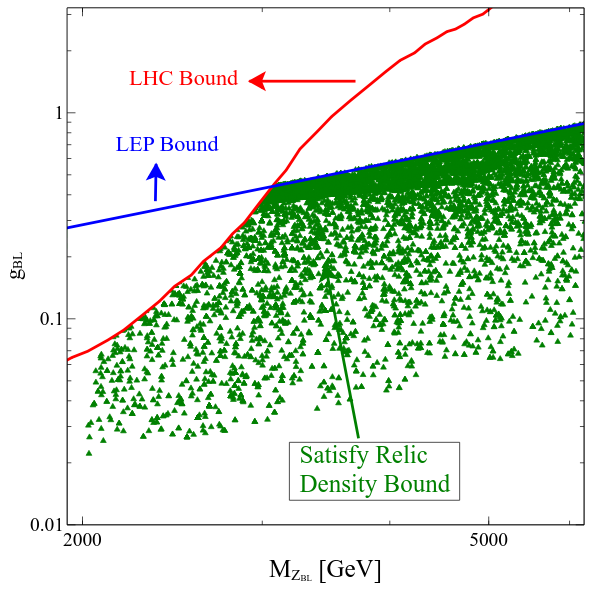}
\includegraphics[angle=0,height=7.50cm,width=8.0cm]{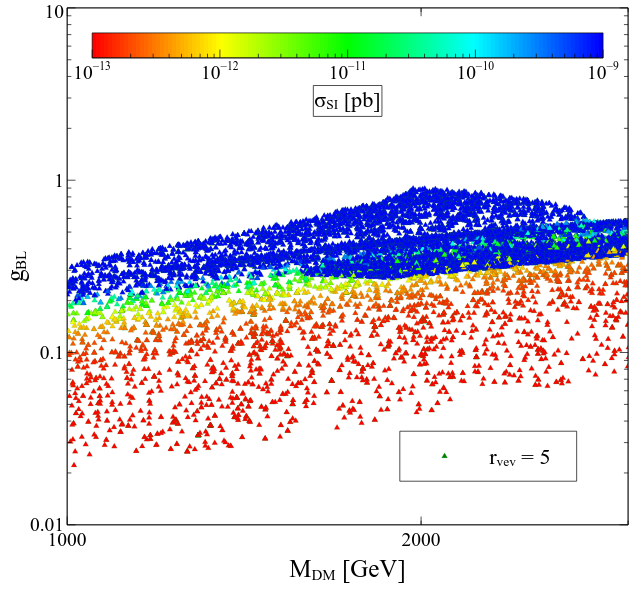}
\caption{Left (Right) Panel : Scatter plot in the $M_{Z_{BL}} - g_{BL}$
($M_{DM} - g_{BL}$) plane after satisfying the DM relic density
bound. The parameters range have been shown in the Table \ref{tab-par-1} and other
parameters have been kept fixed at $r_{{\it vev}} = 5$,
$M_{N_1} = M_{N^{\prime}_1} = 361.4$ GeV,
$M_{N_2} = M_{N^{\prime}_2} = 296.2$ GeV,
$M_{N_3} = M_{N^{\prime}_3} = 111.0$ GeV, 
$M_{h_3} = 250$ GeV, $M_{h_2} = 1000$ GeV,
$h_{e\mu} = 1.58$, $h_{e\tau} = 1.63$, $h_{\mu e} = 2.38$, $h_{\tau e} = 1.34$,
$\sin \beta = 0.07$,
$\Delta M = M_2 - M_1 = M_2-M_{DM} = 50$ GeV.}      
\label{scatt-dm-3}
\end{figure}

In Fig.\,\ref{scatt-dm-3}, we show regions in the
$M_{Z_{BL}}-g_{BL}$ and $M_{DM} - g_{BL}$ planes allowed by the 
DM relic density bound. In the left panel we show the LEP bound
\cite{Carena:2004xs, Cacciapaglia:2006pk, Schael:2013ita} as well as the
LHC dilepton search bound \cite{Chatrchyan:2012oaa, Aad:2014cka, Guo:2015lxa} on the $M_{Z_{BL}}-g_{BL}$ plane\footnote{
LHC, ATLAS and CMS collaborations consider
the Drell-Yan processes ($p\,p \rightarrow \zbl
\rightarrow \bar{l}\,l$, with $l$ = e or $\mu$)
to get the bound in $M_{\zbl}-\gbl$ plane
and they put
lower bound on $M_{\zbl}$ for a particular value of
extra gauge coupling $\gbl$ by searching the dilepton resonance.
For updated bounds
on the mass of extra neutral gauge boson ($Z_{BL}$) at 13
TeV run of LHC,
see Refs.\,\,\cite{Aaboud:2016cth, Aaboud:2016hmk}.} 
{\footnote{LEP consider the processes $e^{+}\,e^{-}
\rightarrow \bar{f}\,f$ ($f \neq e$) above the
Z-pole mass.
They put lower limit on the ratio between the
gauge boson mass and guage coupling by measuring its cross section, which is
$\frac{M_{\zbl}}{g_{BL}} \geq 6-7$ TeV.}.
The region between the red line and the blue line is excluded by the LHC data.
Still, one can see that
a large portion of the area is allowed and can be accessed in the future run of LHC
to test the validity of the present model. On the other hand in the
right panel we show the allowed region in the $M_{DM}-g_{BL}$ plane. Each
point satisfies the DM relic density bound. The color map shows the
corresponding value of the SI direct detection cross-section. 

\section{Conclusion}
\label{conclusion}
In this work we extended the SM by two additional 
gauge groups $\ubl$ and $\umt$. Introducing the $\umt$
gauge group helps us in two way. 
Firstly, it provides a solution to the muon ($g-2$)
anomaly due to the presence of the
extra gauge boson $Z_{\mu\tau}$ and secondly it provides a 
peculiar form to the neutrino mass matrix due to flavour symmetry.
In this work, we generated the light neutrino masses by the inverse
seesaw mechanism. Due to the peculiar form of the neutrino mass
matrix we obtained correlation among the allowed model parameters after
putting constraints on mass squared differences
and mixing angles from the neutrino oscillation data.
In particular, we have shown that 
the parameter values which reproduce neutrino oscillation data for
NH and IH are almost non-overlapping for some of the model parameters.  
However, some parameters are seen to have overlapping
values for both NH and IH. 

We also have studied in detail the DM phenomenology. We have
shown the variation of the DM relic density with its mass
for different values of the other relevant models parameters. 
We have mainly focussed on that portion of the parameter space where
DM dominantly annihilates to the RH neutrinos. 
We have kept $\umt$ gauge boson light in order to
explain the muon ($g-2$) anomaly. Moreover, the small
value of $\gmt$ as required to explain muon
($g-2$) anomaly, makes  $Z_{\mu \tau}$ insignificant
to the cosmic evolution of DM. But the other gauge boson
$Z_{BL}$ which has a TeV scale mass, plays an important
role in DM relic density as well as its direct detection.
Further, we also have two extra Higgs bosons which also
play an important role in the freeze-out processes of DM.
We have found that in our considered mass range for DM,
relic density satisfies the Planck limit
($0.1172\leq\Omega h^2\leq0.1226$) mainly around
the resonance regions of the mediators $h_2$ and $\zbl$
respectively.   
We have explored both the resonance regions separately 
by varying the relevant parameters. We have shown that
near the $h_2$ resonance region a sharp correlation
exists between the DM mass and the mass of the
scalar $h_2$. One could also expect a similar type of
correlation between the DM mass and that of $\ubl$ gauge boson,
but due to the dominance of the scalar mediated diagrams
($h_2$, $h_3$) for the particular values of the gauge
coupling ($g_{BL}$) and gauge boson mass ($M_{Z_{BL}}$),
such correlations are destroyed near the $Z_{BL}$
resonance region. 

In this work, the parameter space have been
chosen in such a way that the dark sector can talk to
the SM particles only via neutral gauge boson $Z_{BL}$ .
Therefore, the direct detection of our DM candidate
$\Sigma_1$ is possible only through the spin independent
elastic scattering mediated by the ${\rm B-L}$
gauge boson $\zbl$.
We have computed the spin independent 
elastic scattering cross section between DM
and nucleon and have compared our results
with the latest exclusion limits obtained from
XENON1T and PandaX-II experiments.
We have found that although some portion of $\sigma_{\rm SI}-M_{DM}$
plane of our present model is already ruled-out
by the present direct detection experiments,
there still remains sufficient region which  
can be tested in the near future by the different
ongoing direct detection experiments like XENON1T ,
PandaX-II and Darwin \cite{Aalbers:2016jon}.
Another test of this model would be via detection at the collider.
One of the signatures will be Drell Yan dilepton production mediated by the
$Z_{BL}$ gauge boson like $pp \rightarrow Z_{BL} \rightarrow l \bar{l}$.
Another interesting search will be di-jet ($2j$) +
missing energy ($\cancel{E}_{T}$) or dilepton ($2l$) +
missing energy ($\cancel{E}_{T}$) by the following processes
\begin{eqnarray}
p p \rightarrow Z_{BL}\,Z_{BL} \rightarrow & 2j (2l) + \cancel{E}_{T}\,.
\end{eqnarray}           
Therefore, the viability of the present model can be tested
both at direct detection as well as collider experiments in near
future.
\section{Acknowledgements}
SK and AB acknowledge the cluster
computing facility at HRI (http://cluster.hri.res.in).
The authors would also like to thank the Department of Atomic Energy
(DAE) Neutrino Project of Harish-Chandra Research Institute.
One of the authors AB acknowledges the financial support from
SERB, Govt. of INDIA through NPDF fellowship (Project No. PDF/2017/000490).
This project has received funding from the European Union's Horizon
2020 research and innovation programme InvisiblesPlus RISE
under the Marie Sklodowska-Curie
grant  agreement  No  690575. This  project  has
received  funding  from  the  European
Union's Horizon  2020  research  and  innovation
programme  Elusives  ITN  under  the Marie
Sklodowska-Curie grant agreement No 674896.

\appendix
\section{Determination of $a_{1}$, $a_{2}$ and $a_3$ matrices}
\label{App:AppendixA}
In studying the DM phenomenology we need to know the value of
the $a_{1,2}$ and $a_3$. Here, we have chosen the
value of the parameters of matrices $m_D$, $\mathcal{M}_{R}$
and $\mu$ which satisfy the neutrino oscillation data for normal
hierarchy as given in the section \ref{iss}. The value of the parameters are as follows,
$Y_e = 0.2219$ GeV, $Y_{\mu} = 0.5619$ GeV, $Y_{\tau} = 1.357$ GeV,
$M_{ee} = 186.15$ GeV, $V_{e\mu} = 124.38$ GeV, $V_{e\tau} = 128.08$ GeV,
$V_{\mu e} = 187.26$ GeV, $M^R_{\mu\mu} = 41.92$ GeV, $M^I_{\mu\mu} = 121.04$ GeV,
 $V_{\tau e} = 104.86$ GeV, $M^R_{\tau\tau} = 72.22$ GeV,
 $M^I_{\tau\tau} = 383.87$ GeV, $Y_{ee} = 1.86\times 10^{-7}$ GeV,
 $Y^R_{\mu\tau} = 2.93\times 10^{-6}$ GeV and $Y^I_{\mu\tau} = 6.14 \times 10^{-7}$ GeV.
 Using these values we find $a_{1,2} \sim m_{D}/(M_{N} \sqrt{2 + \frac{2m_{D}}{M_{N}}})$ and
$a_3 \sim m_{D}/M_{N}$ and carried out our DM analysis.


\begin{thebibliography}{99}
\bibitem{Cowan:1992xc} 
  C.~L.~Cowan, F.~Reines, F.~B.~Harrison, H.~W.~Kruse and A.~D.~McGuire,
  Science {\bf 124}, 103 (1956).


\bibitem{Fukuda:1998mi} 
  Y.~Fukuda {\it et al.} [Super-Kamiokande Collaboration],
  Phys.\ Rev.\ Lett.\  {\bf 81}, 1562 (1998),
  [hep-ex/9807003].


\bibitem{Ahmad:2002jz} 
  Q.~R.~Ahmad {\it et al.} [SNO Collaboration],
  Phys.\ Rev.\ Lett.\  {\bf 89}, 011301 (2002),
  [nucl-ex/0204008].


\bibitem{Eguchi:2002dm} 
  K.~Eguchi {\it et al.} [KamLAND Collaboration],
  Phys.\ Rev.\ Lett.\  {\bf 90}, 021802 (2003),
  [hep-ex/0212021].


\bibitem{An:2015nua} 
  F.~P.~An {\it et al.} [Daya Bay Collaboration],
  Phys.\ Rev.\ Lett.\  {\bf 116}, no. 6, 061801 (2016)
  Erratum: [Phys.\ Rev.\ Lett.\  {\bf 118}, no. 9, 099902 (2017)],
  [arXiv:1508.04233 [hep-ex]].


\bibitem{RENO:2015ksa} 
  J.~H.~Choi {\it et al.} [RENO Collaboration],
  Phys.\ Rev.\ Lett.\  {\bf 116}, no. 21, 211801 (2016),
  [arXiv:1511.05849 [hep-ex]].


\bibitem{Abe:2014bwa} 
  Y.~Abe {\it et al.} [Double Chooz Collaboration],
  JHEP {\bf 1410}, 086 (2014)
  Erratum: [JHEP {\bf 1502}, 074 (2015)],
  [arXiv:1406.7763 [hep-ex]].


\bibitem{Abe:2015awa} 
  K.~Abe {\it et al.} [T2K Collaboration],
  Phys.\ Rev.\ D {\bf 91}, no. 7, 072010 (2015),
  [arXiv:1502.01550 [hep-ex]].


\bibitem{Salzgeber:2015gua} 
  M.~Ravonel Salzgeber [T2K Collaboration],
  arXiv:1508.06153 [hep-ex].


\bibitem{Adamson:2016tbq} 
  P.~Adamson {\it et al.} [NOvA Collaboration],
  Phys.\ Rev.\ Lett.\  {\bf 116}, no. 15, 151806 (2016),
  [arXiv:1601.05022 [hep-ex]].


\bibitem{Adamson:2016xxw} 
  P.~Adamson {\it et al.} [NOvA Collaboration],
  Phys.\ Rev.\ D {\bf 93}, no. 5, 051104 (2016),
  [arXiv:1601.05037 [hep-ex]].


\bibitem{Sakharov:1967dj} 
  A.~D.~Sakharov,
  Pisma Zh.\ Eksp.\ Teor.\ Fiz.\  {\bf 5}, 32 (1967)
  [JETP Lett.\  {\bf 5}, 24 (1967)]
  [Sov.\ Phys.\ Usp.\  {\bf 34}, no. 5, 392 (1991)]
  [Usp.\ Fiz.\ Nauk {\bf 161}, no. 5, 61 (1991)].


\bibitem{Buchmuller:1992qc} 
  W.~Buchmuller and T.~Yanagida,
  Phys.\ Lett.\ B {\bf 302}, 240 (1993).


\bibitem{Buchmuller:1996pa} 
  W.~Buchmuller and M.~Plumacher,
  Phys.\ Lett.\ B {\bf 389}, 73 (1996),
  [hep-ph/9608308].


\bibitem{Dulaney:2010dj} 
  T.~R.~Dulaney, P.~Fileviez Perez and M.~B.~Wise,
  Phys.\ Rev.\ D {\bf 83}, 023520 (2011),
  [arXiv:1005.0617 [hep-ph]].


\bibitem{Sofue:2000jx} 
  Y.~Sofue and V.~Rubin,
  Ann.\ Rev.\ Astron.\ Astrophys.\  {\bf 39}, 137 (2001),
  [astro-ph/0010594].


\bibitem{Bartelmann:1999yn} 
  M.~Bartelmann and P.~Schneider,
  Phys.\ Rept.\  {\bf 340}, 291 (2001),
  [astro-ph/9912508].


\bibitem{Clowe:2003tk} 
  D.~Clowe, A.~Gonzalez and M.~Markevitch,
  Astrophys.\ J.\  {\bf 604}, 596 (2004),
  [astro-ph/0312273].


\bibitem{Biviano:1996bg} 
  A.~Biviano, P.~Katgert, A.~Mazure, M.~Moles, R.~denHartog, J.~Perea and P.~Focardi,
  Astron.\ Astrophys.\  {\bf 321}, 84 (1997),
  [astro-ph/9610168].


\bibitem{Kahlhoefer:2013dca} 
  F.~Kahlhoefer, K.~Schmidt-Hoberg, M.~T.~Frandsen and S.~Sarkar,
  Mon.\ Not.\ Roy.\ Astron.\ Soc.\  {\bf 437}, no. 3, 2865 (2014),
  [arXiv:1308.3419 [astro-ph.CO]].


\bibitem{Harvey:2015hha} 
  D.~Harvey, R.~Massey, T.~Kitching, A.~Taylor and E.~Tittley,
  Science {\bf 347}, 1462 (2015),
  [arXiv:1503.07675 [astro-ph.CO]].


\bibitem{Hinshaw:2012aka} 
  G.~Hinshaw {\it et al.} [WMAP Collaboration],
  Astrophys.\ J.\ Suppl.\  {\bf 208}, 19 (2013),
  [arXiv:1212.5226 [astro-ph.CO]].


\bibitem{Ade:2015xua} 
  P.~A.~R.~Ade {\it et al.} [Planck Collaboration],
  Astron.\ Astrophys.\  {\bf 594}, A13 (2016),
  [arXiv:1502.01589 [astro-ph.CO]].


\bibitem{Khalil:2007dr} 
  S.~Khalil and A.~Masiero,
  Phys.\ Lett.\ B {\bf 665}, 374 (2008),
  [arXiv:0710.3525 [hep-ph]].


\bibitem{FileviezPerez:2010ek} 
  P.~Fileviez Perez and S.~Spinner,
  Phys.\ Rev.\ D {\bf 83}, 035004 (2011),
  [arXiv:1005.4930 [hep-ph]].


\bibitem{Kikuchi:2008xu} 
  T.~Kikuchi and T.~Kubo,
  Phys.\ Lett.\ B {\bf 666}, 262 (2008),
  [arXiv:0804.3933 [hep-ph]].


\bibitem{Fonseca:2011vn} 
  R.~M.~Fonseca, M.~Malinsky, W.~Porod and F.~Staub,
  Nucl.\ Phys.\ B {\bf 854}, 28 (2012),
  [arXiv:1107.2670 [hep-ph]].


\bibitem{Biswas:2017tce} 
  A.~Biswas, S.~Choubey and S.~Khan,
  Eur.\ Phys.\ J.\ C {\bf 77}, no. 12, 875 (2017),
  [arXiv:1704.00819 [hep-ph]].


\bibitem{Biswas:2016yan} 
  A.~Biswas, S.~Choubey and S.~Khan,
  JHEP {\bf 1609}, 147 (2016),
  [arXiv:1608.04194 [hep-ph]].


\bibitem{Mohapatra:1986aw} 
  R.~N.~Mohapatra,
  Phys.\ Rev.\ Lett.\  {\bf 56}, 561 (1986).


\bibitem{Mohapatra:1986bd} 
  R.~N.~Mohapatra and J.~W.~F.~Valle,
  Phys.\ Rev.\ D {\bf 34}, 1642 (1986).


\bibitem{Kang:2006sn} 
  S.~K.~Kang and C.~S.~Kim,
  Phys.\ Lett.\ B {\bf 646}, 248 (2007),
  [hep-ph/0607072].


\bibitem{An:2011uq} 
  H.~An, P.~S.~B.~Dev, Y.~Cai and R.~N.~Mohapatra,
  Phys.\ Rev.\ Lett.\  {\bf 108}, 081806 (2012),
  [arXiv:1110.1366 [hep-ph]].


\bibitem{BhupalDev:2012ru} 
  P.~S.~Bhupal Dev, S.~Mondal, B.~Mukhopadhyaya and S.~Roy,
  JHEP {\bf 1209}, 110 (2012),
  [arXiv:1207.6542 [hep-ph]].


\bibitem{Banerjee:2015gca} 
  S.~Banerjee, P.~S.~B.~Dev, A.~Ibarra, T.~Mandal and M.~Mitra,
  Phys.\ Rev.\ D {\bf 92}, 075002 (2015),
  [arXiv:1503.05491 [hep-ph]].


\bibitem{Dev:2009aw} 
  P.~S.~B.~Dev and R.~N.~Mohapatra,
  Phys.\ Rev.\ D {\bf 81}, 013001 (2010),
  [arXiv:0910.3924 [hep-ph]].

\bibitem{Das:2012ze} 
  A.~Das and N.~Okada,
  Phys.\ Rev.\ D {\bf 88}, 113001 (2013)
  [arXiv:1207.3734 [hep-ph]].

\bibitem{Das:2014jxa} 
  A.~Das, P.~S.~Bhupal Dev and N.~Okada,
  Phys.\ Lett.\ B {\bf 735}, 364 (2014)
  doi:10.1016/j.physletb.2014.06.058
  [arXiv:1405.0177 [hep-ph]].


\bibitem{Mondal:2016kof} 
  S.~Mondal and S.~K.~Rai,
  Phys.\ Rev.\ D {\bf 94}, no. 3, 033008 (2016),
  [arXiv:1605.04508 [hep-ph]].


\bibitem{Banerjee:2013fga} 
  S.~Banerjee, P.~S.~B.~Dev, S.~Mondal, B.~Mukhopadhyaya and S.~Roy,
  JHEP {\bf 1310}, 221 (2013),
  [arXiv:1306.2143 [hep-ph]].


\bibitem{Mondal:2012jv} 
  S.~Mondal, S.~Biswas, P.~Ghosh and S.~Roy,
  JHEP {\bf 1205}, 134 (2012),
  [arXiv:1201.1556 [hep-ph]].


\bibitem{Matsumoto:2010zg} 
  S.~Matsumoto, T.~Nabeshima and K.~Yoshioka,
  JHEP {\bf 1006}, 058 (2010),
  [arXiv:1004.3852 [hep-ph]].


\bibitem{Humbert:2015epa} 
  P.~Humbert, M.~Lindner and J.~Smirnov,
  JHEP {\bf 1506}, 035 (2015),
  [arXiv:1503.03066 [hep-ph]].


\bibitem{Humbert:2015yva} 
  P.~Humbert, M.~Lindner, S.~Patra and J.~Smirnov,
  JHEP {\bf 1509}, 064 (2015),
  [arXiv:1505.07453 [hep-ph]].


\bibitem{Ibarra:2011xn} 
  A.~Ibarra, E.~Molinaro and S.~T.~Petcov,
  Phys.\ Rev.\ D {\bf 84}, 013005 (2011),
  [arXiv:1103.6217 [hep-ph]].


\bibitem{Datta:1992qw} 
  A.~Datta, M.~Guchait and D.~P.~Roy,
  Phys.\ Rev.\ D {\bf 47}, 961 (1993),
  [hep-ph/9208228].


\bibitem{Huitu:2008gf} 
  K.~Huitu, S.~Khalil, H.~Okada and S.~K.~Rai,
  Phys.\ Rev.\ Lett.\  {\bf 101}, 181802 (2008),
  [arXiv:0803.2799 [hep-ph]].


\bibitem{Khalil:2015wua} 
  S.~Khalil and C.~S.~Un,
  Phys.\ Lett.\ B {\bf 763}, 164 (2016),
  [arXiv:1509.05391 [hep-ph]].


\bibitem{Abbas:2015zna} 
  M.~Abbas, S.~Khalil, A.~Rashed and A.~Sil,
  Phys.\ Rev.\ D {\bf 93}, no. 1, 013018 (2016),
  [arXiv:1508.03727 [hep-ph]].


\bibitem{Elsayed:2011de} 
  A.~Elsayed, S.~Khalil and S.~Moretti,
  Phys.\ Lett.\ B {\bf 715}, 208 (2012),
  [arXiv:1106.2130 [hep-ph]].


\bibitem{Khalil:2015naa} 
  S.~Khalil and S.~Moretti,
  Rept.\ Prog.\ Phys.\  {\bf 80}, no. 3, 036201 (2017),
  [arXiv:1503.08162 [hep-ph]].


\bibitem{Abdallah:2015uba} 
  W.~Abdallah, J.~Fiaschi, S.~Khalil and S.~Moretti,
  JHEP {\bf 1602}, 157 (2016),
  [arXiv:1510.06475 [hep-ph]].


\bibitem{Arganda:2015ija} 
  E.~Arganda, M.~J.~Herrero, X.~Marcano and C.~Weiland,
  Phys.\ Lett.\ B {\bf 752}, 46 (2016),
  [arXiv:1508.05074 [hep-ph]].


\bibitem{He:1990pn} 
  X.~G.~He, G.~C.~Joshi, H.~Lew and R.~R.~Volkas,
  Phys.\ Rev.\ D {\bf 43}, 22 (1991).


\bibitem{He:1991qd} 
  X.~G.~He, G.~C.~Joshi, H.~Lew and R.~R.~Volkas,
  Phys.\ Rev.\ D {\bf 44}, 2118 (1991).


\bibitem{Ma:2001md} 
  E.~Ma, D.~P.~Roy and S.~Roy,
  Phys.\ Lett.\ B {\bf 525}, 101 (2002),
  [hep-ph/0110146].


\bibitem{Xing:2015fdg} 
  Z.~z.~Xing and Z.~h.~Zhao,
  Rept.\ Prog.\ Phys.\  {\bf 79}, no. 7, 076201 (2016),
  [arXiv:1512.04207 [hep-ph]].


\bibitem{Biswas:2016yjr} 
  A.~Biswas, S.~Choubey and S.~Khan,
  JHEP {\bf 1702}, 123 (2017),
  [arXiv:1612.03067 [hep-ph]].


\bibitem{Biswas:2017ait} 
  A.~Biswas, S.~Choubey, L.~Covi and S.~Khan,
  JCAP {\bf 1802}, no. 02, 002 (2018),
  [arXiv:1711.00553 [hep-ph]].


\bibitem{Bennett:2004pv} 
  G.~W.~Bennett {\it et al.} [Muon g-2 Collaboration],
  Phys.\ Rev.\ Lett.\  {\bf 92}, 161802 (2004),
  [hep-ex/0401008].


\bibitem{Jegerlehner:2009ry} 
  F.~Jegerlehner and A.~Nyffeler,
  Phys.\ Rept.\  {\bf 477}, 1 (2009),
  [arXiv:0902.3360 [hep-ph]].


\bibitem{Agashe:2014kda} 
  K.~A.~Olive {\it et al.} [Particle Data Group],
  Chin.\ Phys.\ C {\bf 38}, 090001 (2014).

\bibitem{Patra:2016ofq} 
  S.~Patra, W.~Rodejohann and C.~E.~Yaguna,
  JHEP {\bf 1609}, 076 (2016)
  [arXiv:1607.04029 [hep-ph]].
  
  \bibitem{Biswas:2016iyh} 
  A.~Biswas and A.~Gupta,
  JCAP {\bf 1703}, no. 03, 033 (2017)
  Addendum: [JCAP {\bf 1705}, no. 05, A02 (2017)]
  [arXiv:1612.02793 [hep-ph]].
  
  \bibitem{Nanda:2017bmi}
  D.~Nanda and D.~Borah,
  Phys.\ Rev.\ D {\bf 96} (2017) no.11,  115014
  [arXiv:1709.08417 [hep-ph]].

\bibitem{Khalil:2010iu} 
  S.~Khalil,
  Phys.\ Rev.\ D {\bf 82}, 077702 (2010),
  [arXiv:1004.0013 [hep-ph]].


\bibitem{Zhou:2012ds} 
  Y.~L.~Zhou,
  Phys.\ Rev.\ D {\bf 86}, 093011 (2012),
  [arXiv:1205.2303 [hep-ph]].


\bibitem{Law:2013gma} 
  S.~S.~C.~Law and K.~L.~McDonald,
  Phys.\ Rev.\ D {\bf 87}, no. 11, 113003 (2013),
  [arXiv:1303.4887 [hep-ph]].


\bibitem{El-Zant:2013nta} 
  A.~El-Zant, S.~Khalil and A.~Sil,
  Phys.\ Rev.\ D {\bf 91}, no. 3, 035030 (2015),
  [arXiv:1308.0836 [hep-ph]].


\bibitem{Adler:1969gk} 
  S.~L.~Adler,
  Phys.\ Rev.\  {\bf 177}, 2426 (1969).


\bibitem{Bardeen:1969md} 
  W.~A.~Bardeen,
  Phys.\ Rev.\  {\bf 184}, 1848 (1969).


\bibitem{Delbourgo:1972xb} 
  R.~Delbourgo and A.~Salam,
  Phys.\ Lett.\  {\bf 40B}, 381 (1972).


\bibitem{Eguchi:1976db} 
  T.~Eguchi and P.~G.~O.~Freund,
  Phys.\ Rev.\ Lett.\  {\bf 37}, 1251 (1976).


\bibitem{Capozzi:2016rtj} 
  F.~Capozzi, E.~Lisi, A.~Marrone, D.~Montanino and A.~Palazzo,
  Nucl.\ Phys.\ B {\bf 908}, 218 (2016),
  [arXiv:1601.07777 [hep-ph]].


\bibitem{Aprile:2017iyp} 
  E.~Aprile {\it et al.} [XENON Collaboration],
  Phys.\ Rev.\ Lett.\  {\bf 119}, no. 18, 181301 (2017),
  [arXiv:1705.06655 [astro-ph.CO]].


\bibitem{Cui:2017nnn} 
  X.~Cui {\it et al.} [PandaX-II Collaboration],
  Phys.\ Rev.\ Lett.\  {\bf 119}, no. 18, 181302 (2017),
  [arXiv:1708.06917 [astro-ph.CO]].


\bibitem{Belanger:2008sj} 
  G.~Belanger, F.~Boudjema, A.~Pukhov and A.~Semenov,
  Comput.\ Phys.\ Commun.\  {\bf 180}, 747 (2009),
  [arXiv:0803.2360 [hep-ph]].



\bibitem{Geiregat:1990gz} 
  D.~Geiregat {\it et al.} [CHARM-II Collaboration],
  Phys.\ Lett.\ B {\bf 245}, 271 (1990).
  doi:10.1016/0370-2693(90)90146-W

\bibitem{Mishra:1991bv} 
  S.~R.~Mishra {\it et al.} [CCFR Collaboration],
  Phys.\ Rev.\ Lett.\  {\bf 66}, 3117 (1991).
  doi:10.1103/PhysRevLett.66.3117

\bibitem{Altmannshofer:2014pba} 
  W.~Altmannshofer, S.~Gori, M.~Pospelov and I.~Yavin,
  Phys.\ Rev.\ Lett.\  {\bf 113}, 091801 (2014)
  doi:10.1103/PhysRevLett.113.091801
  [arXiv:1406.2332 [hep-ph]].


\bibitem{Edsjo:1997bg} 
  J.~Edsjo and P.~Gondolo,
  Phys.\ Rev.\ D {\bf 56}, 1879 (1997),
  [hep-ph/9704361].


\bibitem{Gondolo:1990dk} 
  P.~Gondolo and G.~Gelmini,
  Nucl.\ Phys.\ B {\bf 360}, 145 (1991).


\bibitem{Belanger:2013oya} 
  G.~Belanger, F.~Boudjema, A.~Pukhov and A.~Semenov,
  Comput.\ Phys.\ Commun.\  {\bf 185}, 960 (2014),
  [arXiv:1305.0237 [hep-ph]].


\bibitem{Alloul:2013bka} 
  A.~Alloul, N.~D.~Christensen, C.~Degrande, C.~Duhr and B.~Fuks,
  Comput.\ Phys.\ Commun.\  {\bf 185}, 2250 (2014),
  [arXiv:1310.1921 [hep-ph]].


\bibitem{Akerib:2016vxi} 
  D.~S.~Akerib {\it et al.} [LUX Collaboration],
  Phys.\ Rev.\ Lett.\  {\bf 118}, no. 2, 021303 (2017),
  [arXiv:1608.07648 [astro-ph.CO]].


\bibitem{Aprile:2015uzo} 
  E.~Aprile {\it et al.} [XENON Collaboration],
  JCAP {\bf 1604}, no. 04, 027 (2016),
  [arXiv:1512.07501 [physics.ins-det]].


\bibitem{Aalbers:2016jon} 
  J.~Aalbers {\it et al.} [DARWIN Collaboration],
  JCAP {\bf 1611}, 017 (2016),
  [arXiv:1606.07001 [astro-ph.IM]].

\bibitem{Carena:2004xs} 
  M.~Carena, A.~Daleo, B.~A.~Dobrescu and T.~M.~P.~Tait,
  Phys.\ Rev.\ D {\bf 70}, 093009 (2004)
  [hep-ph/0408098].

\bibitem{Cacciapaglia:2006pk} 
  G.~Cacciapaglia, C.~Csaki, G.~Marandella and A.~Strumia,
  Phys.\ Rev.\ D {\bf 74}, 033011 (2006)
  [hep-ph/0604111].

\bibitem{Schael:2013ita} 
  S.~Schael {\it et al.} [ALEPH and DELPHI and L3 and OPAL and LEP Electroweak Collaborations],
  Phys.\ Rept.\  {\bf 532}, 119 (2013)
  [arXiv:1302.3415 [hep-ex]].

\bibitem{Chatrchyan:2012oaa} 
  S.~Chatrchyan {\it et al.} [CMS Collaboration],
  Phys.\ Lett.\ B {\bf 720}, 63 (2013)
  [arXiv:1212.6175 [hep-ex]].

\bibitem{Aad:2014cka} 
  G.~Aad {\it et al.} [ATLAS Collaboration],
  Phys.\ Rev.\ D {\bf 90}, no. 5, 052005 (2014)
  [arXiv:1405.4123 [hep-ex]].

\bibitem{Guo:2015lxa} 
  J.~Guo, Z.~Kang, P.~Ko and Y.~Orikasa,
  Phys.\ Rev.\ D {\bf 91}, no. 11, 115017 (2015)
  [arXiv:1502.00508 [hep-ph]].
\bibitem{Aaboud:2016cth} 
  M.~Aaboud {\it et al.} [ATLAS Collaboration],
  Phys.\ Lett.\ B {\bf 761}, 372 (2016)
  [arXiv:1607.03669 [hep-ex]].
\bibitem{Aaboud:2016hmk} 
  M.~Aaboud {\it et al.} [ATLAS Collaboration],
  Eur.\ Phys.\ J.\ C {\bf 76}, no. 10, 541 (2016)
  [arXiv:1607.08079 [hep-ex]].



\end{thebibliography}
\end{document}